\documentclass{acm_proc_article-sp-tweeked}
\usepackage[bookmarks=false]{hyperref}
\usepackage{listings,xcolor}
\usepackage{multicol}
\usepackage[makeroom]{cancel}
\usepackage{color}
\usepackage{subcaption}
\usepackage[scaled=1.04]{couriers}
\usepackage[htt]{hyphenat}
\usepackage{caption}
\begin{document}

\title{Bringing Web Time Travel to MediaWiki}

\subtitle{An Assessment of the Memento MediaWiki Extension}

\numberofauthors{4} 

\author{
Shawn M. Jones, Michael L. Nelson\\
\small Old Dominion University\\
\small Norfolk, VA  23508\\
\small \{sjone,mln\}@cs.odu.edu
\and
Harihar Shankar, Herbert Van de Sompel\\
\small Los Alamos National Laboratory\\
\small Los Alamos, NM, USA\\
\small \{harihar, herbertvg\}@lanl.gov
}

\date{30 May 2014}

\maketitle
\begin{abstract}
We have implemented the Memento MediaWiki Extension Version 2.0, which brings the Memento Protocol to MediaWiki, used by Wikipedia and the Wikimedia Foundation.  Test results show that the extension has a negligible impact on performance. Two 302 status code datetime negotiation patterns, as defined by Memento, have been examined for the extension:  Pattern 1.1, which requires 2 requests, versus Pattern 2.1, which requires 3 requests.  Our test results and mathematical review find that, contrary to intuition, Pattern 2.1 performs better than Pattern 1.1 due to idiosyncrasies in MediaWiki.  In addition to implementing Memento, Version 2.0 allows administrators to choose the optional 200-style datetime negotiation Pattern 1.2 instead of Pattern 2.1.  It also permits administrators the ability to have the Memento MediaWiki Extension return full HTTP 4** and 5** status codes rather than using standard MediaWiki error pages.  Finally, version 2.0 permits administrators to turn off recommended Memento headers if desired.  Seeing as much of our work focuses on producing the correct revision of a wiki page in response to a user's datetime input, we also examine the problem of finding the correct revisions of the embedded resources, including images, stylesheets, and JavaScript; identifying the issues and discussing whether or not MediaWiki must be changed to support this functionality.
\end{abstract}

\section{Introduction}

The Memento Protocol extends HTTP with \emph{datetime negotiation}, a variant of content negotiation.  It uses a given datetime as input and then provides past versions of web pages to a user \cite{memento-rfc:2013}.  These past versions of web pages are referred to as \emph{Mementos}.  It has always been intended for use with both web archives and content management systems (CMS) \cite{vandesompel-ldow10}.  One such CMS is MediaWiki, a common wiki software application famous for its use in Wikipedia.

We discuss the implementation of Memento in the Memento MediaWiki Extension \cite{extension-memento}.

\subsection{Memento Resource Types}

Memento provides several resource types that play a role in datetime negotiation.

The first is the \emph{original resource}, also noted in this paper as a \emph{URI-R}.  It is the page for which we want the past version.  In MediaWiki parlance, it is called a \emph{topic URI}, and refers to the wiki article in its current state.

Then we have the \emph{Memento}, from which the Memento protocol gets its name, also noted in this paper as \emph{URI-M}.  It is the past version of the page.  In MediaWiki parlance, it is called a \emph{oldid page}.

Third, we have the \emph{TimeMap}, also noted in this paper as \emph{URI-T}, which is a resource associated with the original resource from which a list of Mementos for that resource are available.  The TimeMap provides a list of URI-Ms and datetimes in a well-defined format, but does not contain any article content.  TimeMaps have an additional feature in that they can be \emph{paged}, meaning that one TimeMap can lead to others relevant to the same URI-R.

The existing MediaWiki API can construct something akin to a TimeMap.  The API can produce a list of \emph{oldid} and \emph{timestamp} values which correspond to a TimeMap's \emph{memento} URIs and \emph{datetimes}, respectively.  Because the MediaWiki API does not create the Memento URIs, instead only giving \emph{oldids}, an client consuming output from the API will still need to construct URIs to produce the same information as a TimeMap, making such a client unsuitable for the general web.

Listing \ref{lst:example-timemap} shows an example TimeMap produced by the Memento MediaWiki Extension.

Finally, we have the \emph{TimeGate}, noted in this paper as \emph{URI-G}, which is the resource associated with the original resource that provides datetime negotiation.  It is the URI to which the user sends a datetime and receives information about which Memento (URI-M) is closest to it.  The TimeGate only processes and redirects; it provides no representations itself.  There is no MediaWiki equivalent to the TimeGate.  The Memento MediaWiki Extension provides this functionality.

Table \ref{tab:examples} provides example URIs that correspond to each of these resource types once the Memento MediaWiki Extension is installed.

\begin{table*}[t]
\begin{center}
\small
	\begin{tabular}{| l | l | l |}
	\hline
	\textbf{Memento} & \textbf{Memento} & \textbf{Example} \\
	\textbf{Resource Type} & \textbf{Resource Notation} & \\
	\hline
	Original Resource & URI-R & http://ws-dl-05.cs.odu.edu/demo/index.php/Daenerys\_Targaryen \\
	\hline
	Memento & URI-M & http://ws-dl-05.cs.odu.edu/demo/index.php?title=Daenerys\_Targaryen\&oldid=27870 \\
	\hline
	TimeGate & URI-G & http://ws-dl-05.cs.odu.edu/demo/index.php/Special:TimeGate/Daenerys\_Targaryen \\
	\hline 
	TimeMap & URI-T & http://ws-dl-05.cs.odu.edu/demo/index.php/Special:TimeMap/Daenerys\_Targaryen \\
	\hline
	\end{tabular}
\end{center}
\caption{Examples of Memento Resources From the Memento MediaWiki Extension}
\label{tab:examples}
\end{table*}

\begin{lstlisting}[frame=single,basicstyle=\scriptsize\ttfamily,captionpos=b, breaklines=true, caption={Example TimeMap from the Memento MediaWiki Extension}, label={lst:example-timemap}, float=*]
<http://ws-dl-05.cs.odu.edu/demo/index.php/Special:TimeMap/20130711203756/-1/Daenerys_Targaryen>; rel="self"; type="application/link-format"; from="Sat, 23 Feb 2013 01:55:23 GMT"; until="Thu, 11 Jul 2013 20:36:08 GMT",
<http://ws-dl-05.cs.odu.edu/demo/index.php/Special:TimeMap/20130223015523/-1/Daenerys_Targaryen>; rel="timemap"; type="application/link-format";from="Wed, 19 Sep 2012 16:23:26 GMT"; until="Sat, 02 Feb 2013 01:18:43 GMT",
<http://ws-dl-05.cs.odu.edu/demo/index.php/Special:TimeMap/20130711203608/1/Daenerys_Targaryen>; rel="timemap"; type="application/link-format";from="Thu, 11 Jul 2013 20:37:56 GMT"; until="Fri, 27 Sep 2013 20:48:24 GMT",
<http://ws-dl-05.cs.odu.edu/demo/index.php/Special:TimeGate/Daenerys_Targaryen>; rel="timegate",
<http://ws-dl-05.cs.odu.edu/demo/index.php/Daenerys_Targaryen>; rel="original latest-version",
<http://ws-dl-05.cs.odu.edu/demo/index.php?title=Daenerys_Targaryen&oldid=90020>; rel="memento"; datetime="Sat, 23 Feb 2013 01:55:23 GMT",
<http://ws-dl-05.cs.odu.edu/demo/index.php?title=Daenerys_Targaryen&oldid=91783>; rel="memento"; datetime="Wed, 13 Mar 2013 16:22:23 GMT",
<http://ws-dl-05.cs.odu.edu/demo/index.php?title=Daenerys_Targaryen&oldid=93106>; rel="memento"; datetime="Fri, 29 Mar 2013 23:25:08 GMT",
<http://ws-dl-05.cs.odu.edu/demo/index.php?title=Daenerys_Targaryen&oldid=93753>; rel="memento"; datetime="Thu, 11 Apr 2013 01:55:33 GMT",
<http://ws-dl-05.cs.odu.edu/demo/index.php?title=Daenerys_Targaryen&oldid=94427>; rel="memento"; datetime="Thu, 25 Apr 2013 05:30:44 GMT",
<http://ws-dl-05.cs.odu.edu/demo/index.php?title=Daenerys_Targaryen&oldid=94605>; rel="memento"; datetime="Fri, 26 Apr 2013 16:52:08 GMT",
<http://ws-dl-05.cs.odu.edu/demo/index.php?title=Daenerys_Targaryen&oldid=95821>; rel="memento"; datetime="Tue, 07 May 2013 19:30:38 GMT",
<http://ws-dl-05.cs.odu.edu/demo/index.php?title=Daenerys_Targaryen&oldid=95824>; rel="memento"; datetime="Tue, 07 May 2013 19:40:25 GMT",
...
\end{lstlisting}

\subsection{Structure of this Paper}
In this paper, we discuss the architecture of the extension, how it was designed to support Memento while also addressing Wikimedia's concerns, and discuss a TimeGate design choice specific to this effort.  Then we detail the configuration options for the extension.  After that, we use experimental data to show that the Memento MediaWiki Extension does not negatively affect performance.  Finally, we detail the work that remains to be done, jointly with the MediaWiki team, to bring full time travel capability to MediaWiki.

\section{Prior Work}

Additional interest exists in providing time travel capability to MediaWiki, as is evidenced by the Time Machine Extension \cite{extension-timemachine}, and the BackwardsTimeTravel Extension \cite{extension-backwardstimetravel}.  While these extensions do provide the ability to view previous versions of pages, they do not support the Memento protocol that specifies an interoperable approach for temporal access to resource versions, which is meanwhile supported by all major public web archives, worldwide.

These extensions also do not follow the RESTful principle in identifying additional resources for the client to consume \cite{rest}, whereas the Memento MediaWiki Extension, in compliance with the Memento protocol, applies common ``follow your nose'' techniques to lead clients to TimeGates, TimeMaps, and additional Mementos.

Parsoid \cite{extension-parsoid} offers the ability to turn MediaWiki syntax into HTML documents while also attempting to preserve images, stylesheets, and other embedded content.  It does not support Memento, and does not provide real-time access to all of the revisions of a MediaWiki page.

The Collection extension \cite{extension-collection}, is used to preserve wiki pages, with the intent of rending them with mwlib \cite{mwlib} and preserving them in book form for physical reproduction with a service like PediaPress \cite{pediapress}.  This extension only works with the version of the page captured when the book is created by a user.  It is a form of on-demand web archiving, but does not support Memento.

Vi{\'e}gas, Wattenberg, and Dave \cite{viegas} detail the use of \emph{History Flow}, a visualization tool that allows a user to view broad trends in revision histories.  \emph{History Flow} is useful for performing analysis on MediaWiki edits, but does not allow a user to browse past versions of a wiki, nor does it support Memento.

One could manually perform datetime negotiation using MediaWiki's history pages, but this is very time consuming for the individual.

As noted above, one could use the MediaWiki API to perform the functions of Memento, but only a MediaWiki-aware client could construct URIs from the data returned from the API.  Memento provides a web standard way of accessing previous versions of web resources.

Finally, one could use one of the many public web archives to browse past revisions of MediaWiki content.  For this to be effective, the web archive must already be archiving the content of a MediaWiki installation.  Even if a web archive is archiving the content, they will likely not have access to every revision of a given page, making MediaWiki's native access to this data superior for those seeking to view every last past revision of an article.  For example, the page \url{http://awoiaf.westeros.org/index.php/Jaime_Lannister} has 29 revisions between May 2009 and December 2009, but the Internet Archive contains a revision from April 15, 2009 followed by another on March 27, 2010, missing the 29 revisions from the rest of 2009.

\section{Design and Architecture}

MediaWiki provides a utility called a \texttt{SpecialPage} to perform specific functions not covered otherwise.  When creating an extension, one may use these \texttt{SpecialPages} to centralize additional functionality, if necessary.

\begin{figure}[h!]
\centering
	\includegraphics[width=0.3\textwidth]{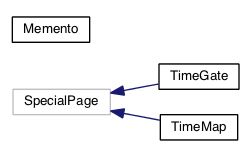}
\caption{Version 1.0 Class Hierarchy Diagram}
\label{fig:version1-classHierarchy}
\end{figure}

Version 1.0 of the Memento MediaWiki Extension used SpecialPages for both the TimeGate and TimeMap implementations.  Beyond this, all logic was embedded in functions.

The class hierarchy diagram for Version 1.0 is shown in Figure \ref{fig:version1-classHierarchy}.

The original extension implemented Pattern 2.1 (i.e., 302 response where URI-R$\neq$URI-G and distinct URI-M) of RFC 7089.

After much discussion with the Wikimedia team \cite{bugzilla}, it was determined that additional work was needed to ready the code for use by Wikipedia.  A generalized list indicates that the extension needed to:
\begin{enumerate}
\item follow MediaWiki coding conventions \cite{conventions}
\item follow MediaWiki's PHP coding conventions \cite{php-conventions}
\item follow the Security checklist for developers \cite{security-checklist}
\item follow MediaWiki's standards for Extension development \cite{writing-extension}
\item not require changes to core MediaWiki code
\item control the use of global variables, avoiding them if possible
\item work with the latest version of MediaWiki
\item avoid injection vulnerabilities when building links
\item limit expensive database operations where possible
\item avoid using deprecated MediaWiki code
\item improve code quality
\end{enumerate}

\begin{figure*}

\begin{subfigure}{1.0\textwidth}
\centering
	\includegraphics[width=0.5\textwidth]{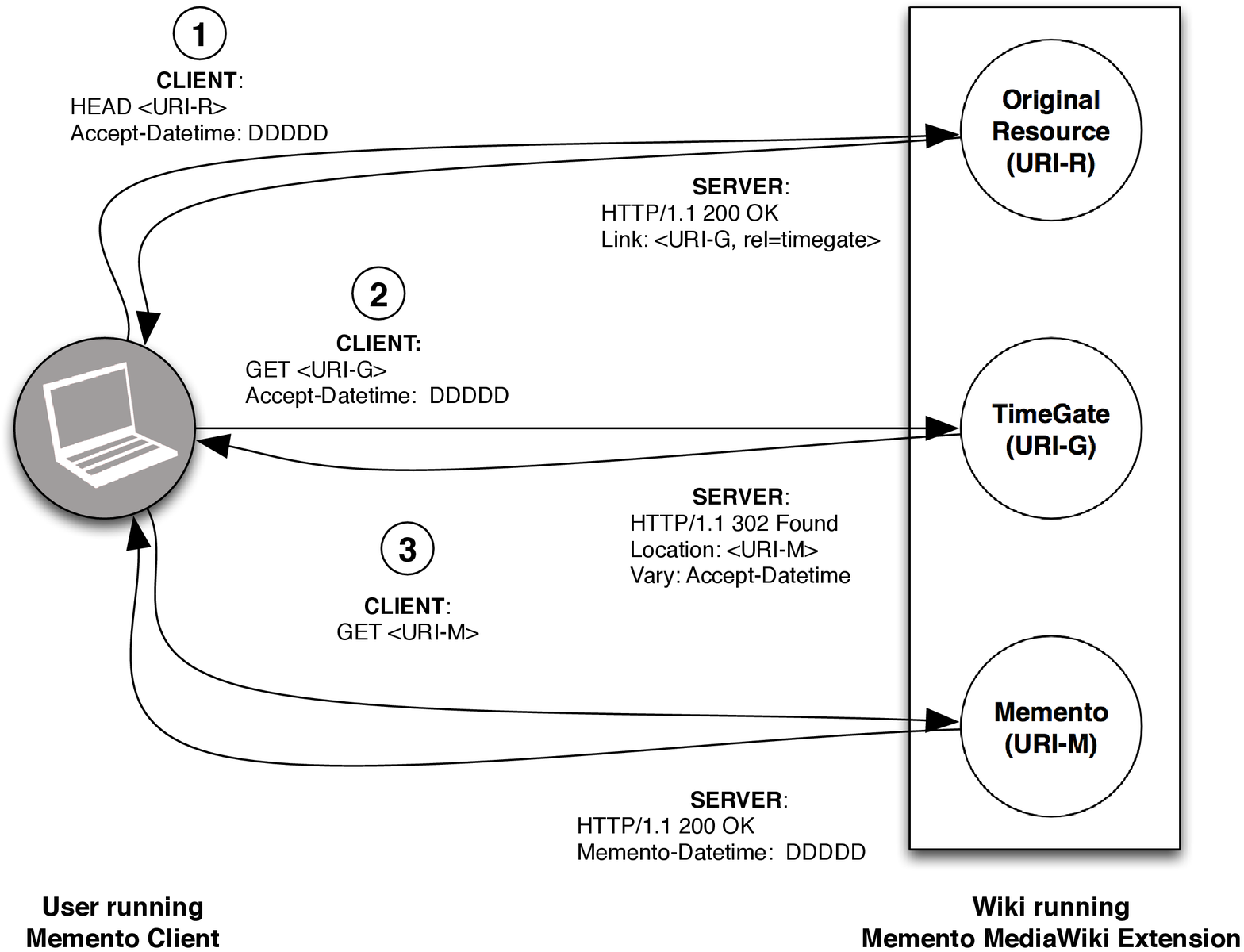}
\caption{Memento Pattern 2.1 Overview with Only Salient Headers, Methods, and Responses}
\label{fig:salient-pattern2.1}

\end{subfigure}

\begin{subfigure}{1.0\textwidth}

\begin{lstlisting}[frame=single,basicstyle=\scriptsize\ttfamily,captionpos=b, breaklines=true, caption={Memento MediaWiki Extension Example Response for step 1 (URI-R) of Memento Pattern 2.1 (Memento headers in red)}, label={lst:example-response-pattern2.1-step1}]
HTTP/1.1 200 OK
Date: Sun, 25 May 2014 21:39:02 GMT
Server: Apache
X-Content-Type-Options: nosniff
Link: ~{\color{red} <http://ws-dl-05.cs.odu.edu/demo/index.php/Daenerys\_Targaryen>;~ ~{\color{red}rel="original~ ~{\color{red}latest-version",<http://ws-dl-05.cs.odu.edu/demo/index.php/Special:TimeGate/Daenerys\_Targaryen>;~ ~{\color{red} rel="timegate",<http://ws-dl-05.cs.odu.edu/demo/index.php/Special:TimeMap/Daenerys\_Targaryen>;~ ~{\color{red} rel="timemap";~ ~{\color{red} type="application/link-format"~
Content-language: en
Vary: Accept-Encoding,Cookie
Cache-Control: s-maxage=18000, must-revalidate, max-age=0
Last-Modified: Sat, 17 May 2014 16:48:28 GMT
Connection: close
Content-Type: text/html; charset=UTF-8
\end{lstlisting}

\end{subfigure}

\begin{subfigure}{1.0\textwidth}

\begin{lstlisting}[frame=single,basicstyle=\scriptsize\ttfamily,captionpos=b, breaklines=true, caption={Memento MediaWiki Extension Example Response for step 2 (URI-G) of Memento Pattern 2.1 (Memento headers in red)}, label={lst:example-response-pattern2.1-step2}]
HTTP/1.1 302 Found
Date: Sun, 25 May 2014 21:43:08 GMT
Server: Apache
X-Content-Type-Options: nosniff
Vary: Accept-Encoding,~{\color{red}Accept-Datetime~
Location: ~{\color{red} http://ws-dl-05.cs.odu.edu/demo/index.php?title=Daenerys\_Targaryen\&oldid=1499~
Link: ~{\color{red} <http://ws-dl-05.cs.odu.edu/demo/index.php/Special:TimeMap/Daenerys\_Targaryen>;~ ~{\color{red} rel="timemap";~ ~{\color{red} type="application/link-format",<http://ws-dl-05.cs.odu.edu/demo/index.php/Daenerys\_Targaryen>;~ ~{\color{red} rel="original~ ~{\color{red} latest-version"~
Connection: close
Content-Type: text/html; charset=UTF-8
\end{lstlisting}

\end{subfigure}

\begin{subfigure}{1.0\textwidth}

\begin{lstlisting}[frame=single,basicstyle=\scriptsize\ttfamily,captionpos=b, breaklines=true, caption={Memento MediaWiki Extension Example Response for step 3 (URI-M) of Memento Pattern 2.1 (Memento headers in red)}, label={lst:example-response-pattern2.1-step3}]
HTTP/1.1 200 OK
Date: Sun, 25 May 2014 21:46:12 GMT
Server: Apache
X-Content-Type-Options: nosniff
~{\color{red}Memento-Datetime: Sun, 22 Apr 2007 15:01:20 GMT~
Link: ~{\color{red} <http://ws-dl-05.cs.odu.edu/demo/index.php/Daenerys\_Targaryen>;~ ~{\color{red}rel="original~ ~{\color{red} latest-version",~ ~{\color{red} <http://ws-dl-05.cs.odu.edu/demo/index.php/Special:TimeGate/Daenerys\_Targaryen>;~ ~{\color{red}  rel="timegate",~ ~{\color{red} <http://ws-dl-05.cs.odu.edu/demo/index.php/Special:TimeMap/Daenerys\_Targaryen>;~ ~{\color{red}rel="timemap";~ ~{\color{red} type="application/link-format"~
Content-language: en
Vary: Accept-Encoding,Cookie
Expires: Thu, 01 Jan 1970 00:00:00 GMT
Cache-Control: private, must-revalidate, max-age=0
Connection: close
Content-Type: text/html; charset=UTF-8
\end{lstlisting}

\end{subfigure}

\caption{Diagram and Example Response Headers for Memento Pattern 2.1}

\end{figure*}

\begin{figure*}

\begin{subfigure}{1.0\textwidth}
\centering
	\includegraphics[width=0.5\textwidth]{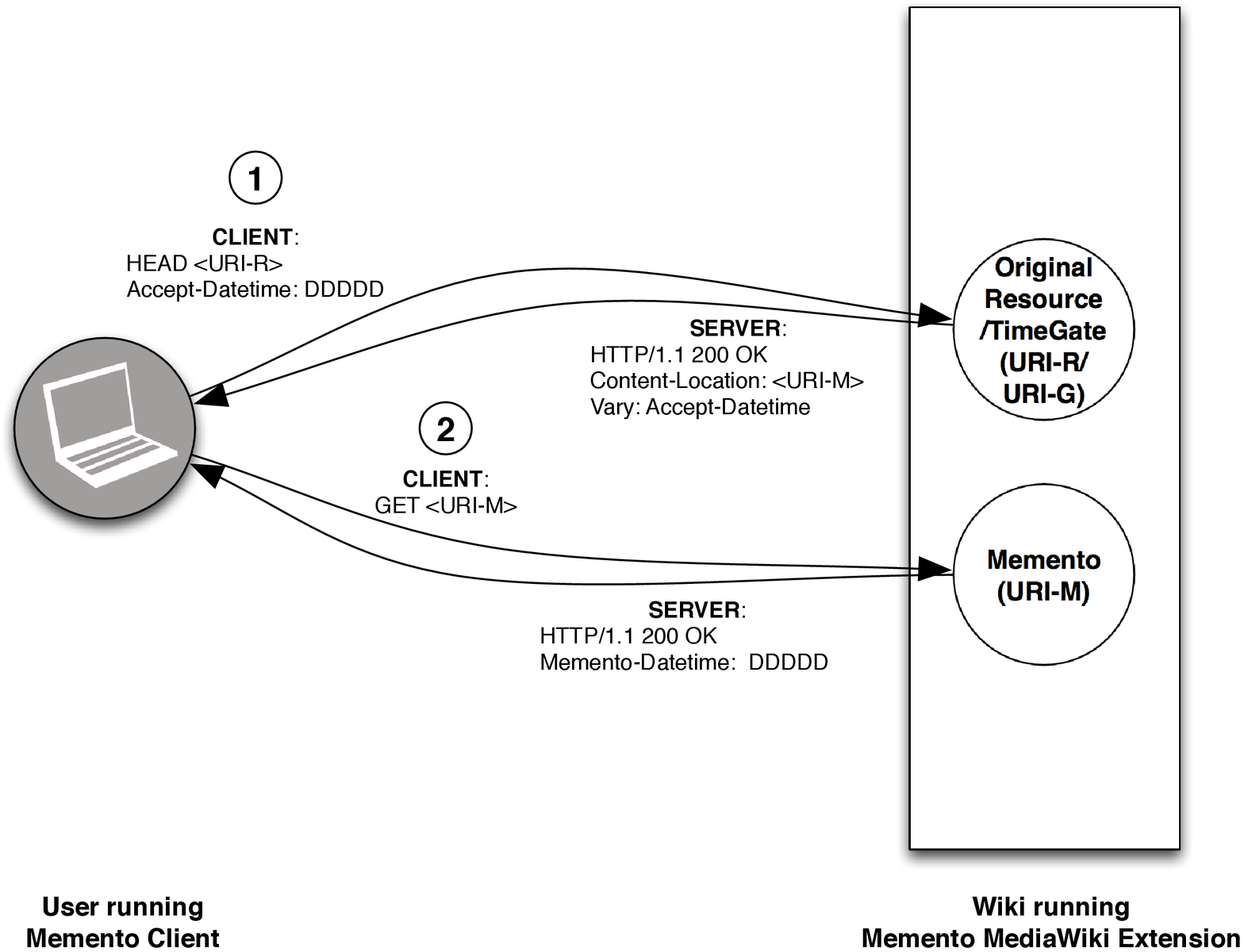}
\caption{Memento Pattern 1.2 Overview with Only Salient Headers, Methods, and Responses}
\label{fig:salient-pattern1.2}
\end{subfigure}

\begin{subfigure}{1.0\textwidth}
\begin{lstlisting}[frame=single,basicstyle=\scriptsize\ttfamily,captionpos=b, breaklines=true, caption={Memento MediaWiki Extension Example Response for step 1 (URI-R=URI-G) of Memento Pattern 1.2 (Memento headers in red)}, label={lst:example-response-pattern-1.2-step1}]
HTTP/1.1 200 OK
Date: Wed, 14 May 2014 16:03:27 GMT
Server: Apache/2.2.15 (Red Hat)
X-Powered-By: PHP/5.3.3
X-Content-Type-Options: nosniff
~{\color{red}Memento-Datetime: Sun, 22 Apr 2007 15:01:20 GMT~
Content-Location: ~{\color{red}http://ws-dl-05.cs.odu.edu/demo-200-style/index.php?title=Daenerys\_Targaryen\&oldid=1499~
Link: ~{\color{red}<http://ws-dl-05.cs.odu.edu/demo-200-style/index.php/Special:TimeMap/Daenerys\_Targaryen>;~ ~{\color{red}rel="timemap";~ ~{\color{red} type="application/link-format",<http://ws-dl-05.cs.odu.edu/demo-200-style/index.php/Daenerys\_Targaryen>;~ ~{\color{red} rel="original~ ~{\color{red} latest-version~ ~{\color{red}timegate"~
Content-language: en
Vary: Accept-Encoding,~{\color{red}Accept-Datetime~,Cookie
Cache-Control: s-maxage=18000, must-revalidate, max-age=0
Last-Modified: Sat, 22 Mar 2014 02:47:12 GMT
Connection: close
Content-Type: text/html; charset=UTF-8
\end{lstlisting}

\end{subfigure}

\begin{subfigure}{1.0\textwidth}

\begin{lstlisting}[frame=single,basicstyle=\scriptsize\ttfamily,captionpos=b, breaklines=true, caption={Memento MediaWiki Extension Example Response for step 2 (URI-M) of Memento Pattern 1.2 (Memento headers in red)}, label={lst:example-response-pattern-1.2-step2}]
HTTP/1.1 200 OK
Date: Wed, 14 May 2014 16:07:34 GMT
Server: Apache/2.2.15 (Red Hat)
X-Powered-By: PHP/5.3.3
X-Content-Type-Options: nosniff
~{\color{red}Memento-Datetime: Sun, 22 Apr 2007 15:01:20 GMT~
Link: ~{\color{red}<http://ws-dl-05.cs.odu.edu/demo-200-style/index.php/Daenerys\_Targaryen>;~ ~{\color{red}rel="original~ ~{\color{red} latest-version~ ~{\color{red} timegate",~ ~{\color{red}<http://ws-dl-05.cs.odu.edu/demo-200-style/index.php/Special:TimeMap/Daenerys\_Targaryen>;~ ~{\color{red}rel="timemap";~ ~{\color{red} type="application/link-format"~
Content-language: en
Vary: Accept-Encoding,Cookie
Expires: Thu, 01 Jan 1970 00:00:00 GMT
Cache-Control: private, must-revalidate, max-age=0
Connection: close
Content-Type: text/html; charset=UTF-8
\end{lstlisting}

\end{subfigure}

\caption{Diagram and Example Response Headers for Memento Pattern 1.2}

\end{figure*}

Thus, version 2.0 was started to address these issues in hopes that it would be acceptable to the Wikimedia community.  In addition, the following new features were to be added:
\begin{itemize}
\item in addition to Pattern 2.1 (i.e., 302 response where URI-R$\neq$URI-G and distinct URI-M) shown in Figure \ref{fig:salient-pattern2.1} and Listings \ref{lst:example-response-pattern2.1-step1}, \ref{lst:example-response-pattern2.1-step2}, and \ref{lst:example-response-pattern2.1-step3}; the MediaWiki administrator may opt instead to use RFC 7089 Pattern 1.2 (i.e., 200 response where URI-R=URI-G and distinct URI-M) shown in Figure \ref{fig:salient-pattern1.2} and Listings \ref{lst:example-response-pattern-1.2-step1} and \ref{lst:example-response-pattern-1.2-step2}
\item allow the MediaWiki administrator the option of disabling \emph{recommended} Memento Link header relations to save on performance
\item allow the MediaWiki administrator the option of choosing between actual HTTP status codes or MediaWiki-compliant 200 responses containing error messages in the entity body (e.g., ``soft-404'' responses \cite{bar-yossef})
\end{itemize}

\begin{figure}[h!]
\centering
	\includegraphics[width=0.5\textwidth]{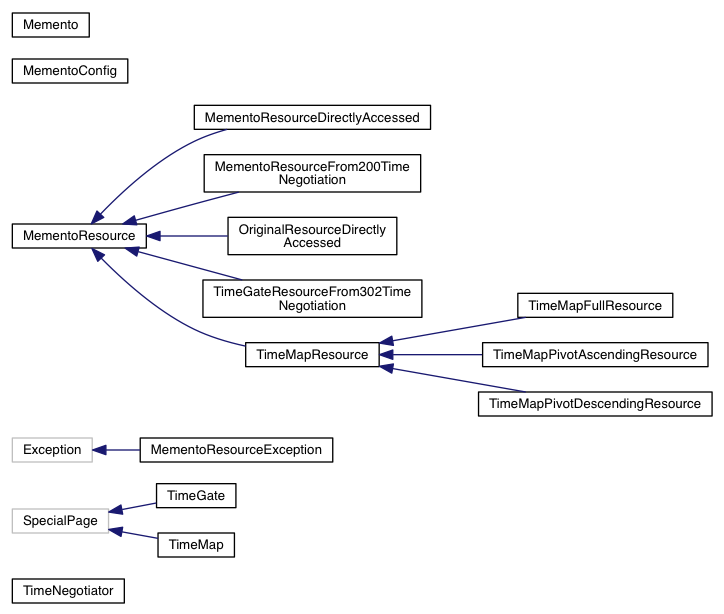}
\caption{Version 2.0 Class Hierarchy Diagram}	
\label{fig:version2-classHierarchy}
\end{figure}

Figure \ref{fig:version2-classHierarchy} shows the improved architecture of the Memento MediaWiki extension to address these concerns and new features.

Version 2.0 of the Memento MediaWiki Extension partitioned functionality into individual classes so that MediaWiki's objects and functions could be consumed and utilized more efficiently, increasing performance while also addressing many of the concerns from the list above.

The \texttt{Memento} class is the extension entry point for URI-R and URI-M work, implementing a \emph{Mediator} design pattern \cite{design-patterns}.  It uses the \texttt{BeforeParserFetchTempateAndtitle} hook \cite{beforeparserfetchtemplateandtitle} to ensure that the revision of an embedded article template matches the revision of the wiki article.  It uses the \texttt{ArticleViewHeader} hook \cite{articleviewheader} to insert Memento headers into the responses.  Finally, it uses the \texttt{BeforePageDisplay} hook \cite{beforepagedisplay} to change the entity body of a page for Pattern 1.2 responses.

Global variables are controlled using the \texttt{MementoConfig} class.  This way all extension configuration options (controlled as globals, as is the MediaWiki convention) are read and stored in one place in a controlled fashion.  All other use of global variables have been removed from the code by using MediaWiki's native functions as much as possible.

As shown in Table \ref{table:classes} the \texttt{MementoResource} family of classes implement the different resource types used in the Memento framework.  This architecture was chosen to improve code quality, while also supporting code extension and reusability. These classes, with the exception of \texttt{TimeGateResourceFrom302TimeNegotiation}, are selected based on the HTTP request using a \emph{Factory Method}.  This \emph{Factory Method}, combined with a \emph{Strategy} pattern, and utilizing \emph{Template Methods}, makes sure the framework is easily extendable to include additional future patterns and resource types.

TimeMaps can be paged, allowing a machine client to follow one TimeMap to another and another using the ``follow your nose'' principle of REST.  TimeMap URIs are constructed by the Memento MediaWiki Extension as shown in the examples in Table \ref{tab:timemapex}.  Arguments, specified as part of the URI, indicate which TimeMaps should be returned.  A \texttt{/-1/} following a datetime in the URI indicates that a TimeMap containing mementos prior to that datetime should be returned.  A \texttt{/1/} following a datetime in the URI indicates that a TimeMap containing mementos after that datetime should be returned.  A URI containing no datetime returns the latest Mementos for the given wiki article and a link to the next TimeMap, if there are more than 500 Mementos.

The \texttt{TimeMap} SpecialPage class also uses this same combination of design patterns to act according to how it are called.  For, example, if the \texttt{TimeMap} SpecialPage is called using a \texttt{/-1/} following a datetime in the URI, then a \newline\texttt{TimeMapPivotDescendingResource} object is instantiated to provide paged TimeMaps below the given datetime.  Likewise a \texttt{/1/} following a datetime in the URI instantiates a \texttt{TimeMapPivotAscendingResource} object, providing paged TimeMaps above the given datetime.  If no pivot is given in the URI, then a \texttt{TimeMapFullResource} object is instantiated, giving the full first page of the TimeMap from the current date.

The \texttt{TimeNegotiator} centralizes all time negotiation functionality.  This way time negotiation is performed using the same algorithm, whether we are using Pattern 1.2 or Pattern 2.1.

\begin{table*}
\begin{center}
\scriptsize
	\begin{tabular}{| l | l |}
	\hline
	\textbf{Meaning} & \textbf{TimeGate URI} \\
	\hline
	Get TimeMap for the & http://ws-dl-05.cs.odu.edu/demo/index.php/Special:TimeMap/Daenerys\_Targaryen \\
	latest 500 Mementos & \\
	for the wiki article & \\
	``Daenerys Targaryen'' & \\
	\hline
	Get TimeMap for the & http://ws-dl-05.cs.odu.edu/demo/index.php/Special:TimeMap/20110630000000/-1/Daenerys\_Targaryen \\
	500 Mementos (or less) & \\
	prior to June 30, 2011 & \\
	at midnight & \\
	\hline
	Get TimeMap for the & http://ws-dl-05.cs.odu.edu/demo/index.php/Special:TimeMap/20110630000000/1/Daenerys\_Targaryen \\
	500 Mementos (or less) & \\
	after June 30, 2011 & \\
	at midnight & \\
	\hline
	\end{tabular}
\end{center}
\caption{Examples of TimeMap URIs From the Memento MediaWiki Extension}
\label{tab:timemapex}
\end{table*}

\begin{table}
\small
\begin{center}
	\begin{tabular}{ | l | l |}
	\hline
	\textbf{Extension Class} & \textbf{Memento} \\
	& \textbf{Resource} \\
	& \textbf{Type} \\ \hline
	MementoResourceDirectlyAccessed & URI-M \\ \hline
	MementoResourceFrom200TimeNegoation & URI-R \\
	& URI-M \\ 
	& URI-G \\
	& (Pattern 1.2) \\ \hline
	OriginalResourceDirectlyAccessed & URI-R \\ \hline
	TimeGateResourceFrom302TimeNegotiation & URI-G \\
	& (Pattern 2.1) \\ \hline
	TimeMapResource (class family): & URI-T \\ 
	\hspace{1em} TimeMapFullResource & \\
	\hspace{1em} TimeMapPivotAscendingResource & \\
	\hspace{1em} TimeMapPivotDescendingResource & \\ \hline
	\end{tabular}
\end{center}
\caption{Version 2.0 Memento MediaWiki Extension \texttt{MementoResource} Class Family Members Mapped To Their Memento Resource Type}
\label{table:classes}
\end{table}

Once this architecture was in place, we were able to address lingering design decisions.

\subsection{TimeGate Design Decision}

In addition to implementing Pattern 1.2, two possible Time\hyp Gate design options were reviewed to determine which would be best suited to be the default pattern in the Memento MediaWiki Extension \cite{blog-timegate-design}.

We evaluated the use of Pattern 1.1 and Pattern 2.1 from RFC 7089.  Both patterns require a Memento client to find the URI-G from header information in the URI-R response.

Pattern 2.1 uses distinct URIs for URI-R and URI-G.  Figure \ref{fig:salient-pattern2.1} shows a simplified diagram of a Pattern 2.1 exchange.

Pattern 1.1 uses the same URI for both URI-R and URI-G, allowing a resource to function as its own TimeGate, meaning that the client can short-circuit the process by one request.

Version 1.0 of the Memento MediaWiki Extension utilized Pattern 2.1, but Pattern 1.1 was explored to save on network traffic and improve performance.

As can be seen in Figure \ref{fig:salient-pattern2.1}, Pattern 2.1 requires three request-response pairs to retrieve a Memento.
\begin{align} \label{eq:pat2.1}
d_{p2.1} &= a + RTT_a + b + RTT_b + M + RTT_M
\end{align}
Equation \ref{eq:pat2.1} calculates the duration of using Pattern 2.1, where $a$ is time the Memento MediaWiki Extension takes to generate the URI-R response in step 1, $b$ is the time it takes to generate the URI-G response in step 2, and $M$ is the time it takes to generate the URI-M response in step 3.  $RTT_a$, $RTT_b$, and $RTT_M$ is defined as \emph{round-trip-time}, which is ``the time it takes for a small packet to travel from client to server and then back to the client'' \cite{networking-book}, for transmitting the data computed during $a$, $b$, and $M$.

\setlength{\abovecaptionskip}{15pt plus 3pt minus 2pt}

\begin{figure}[h!]
\centering
	\includegraphics[width=0.5\textwidth]{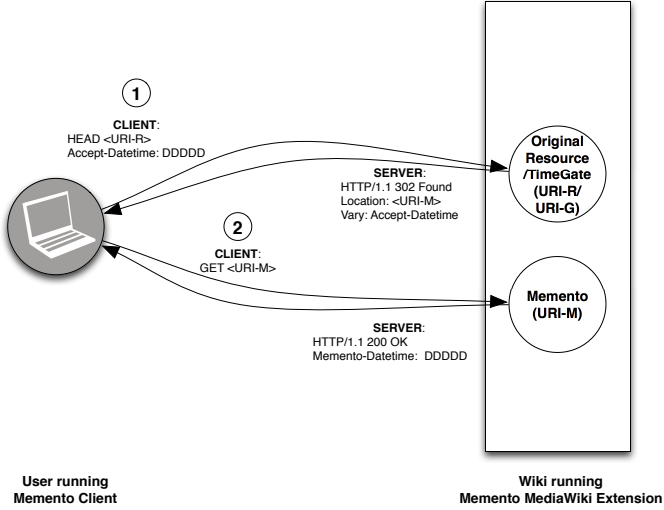}
\caption{Memento Pattern 1.1 Overview with Only Salient Headers, Methods, and Responses; note the 302 response for step 1 that differentiates it from Figure \ref{fig:salient-pattern1.2}}
\label{fig:salient-pattern1.1}
\end{figure}

\setlength{\abovecaptionskip}{10pt plus 0pt minus 0pt}

Figure \ref{fig:salient-pattern1.1} shows a simplified diagram of Pattern 1.1, which requires two request-response pairs to retrieve a Memento.
\begin{align} \label{eq:pat1.1}
d_{p1.1} &= B + RTT_B + M + RTT_M
\end{align}
Equation \ref{eq:pat1.1} calculates the duration for using Pattern 1.1, where $B$ is the time it takes to generate the URI-G response in step 1.  Just like in Equation \ref{eq:pat2.1}, $M$ and $RTT_M$ are the same.  The term $RTT_B$ is the round-trip time to receive and transmit the results of the calculation done during $B$.

Our intuition was that Pattern 1.1 should be faster.  It has fewer round trips to make between the client and server.

For Pattern 1.1 to be the better choice for performance, $d_{p1.1} \textless d_{p1.2}$, which leads to Equation \ref{eq:relationshipstart}.
\begin{align} \label{eq:relationshipstart}
d_{p1.1} & \textless d_{p1.2} \nonumber \\
B + RTT_B + \bcancel{M} + \bcancel{RTT_M} \nonumber & \\
& < \nonumber \\
& a + RTT_a + \nonumber \\
& b + RTT_b + \bcancel{M} + \bcancel{RTT_M } \nonumber \\
B + RTT_B & < a + RTT_a + b + RTT_b
\end{align}

TimeGate responses consist of 302 status messages in response to a GET request.  The difference between the number of bytes in a request and response conversation should differ only by a few bytes at most between Pattern 1.1 and 2.1.  If we consider that a TimeGate response will be equivalent regardless of pattern implemented, then $RTT_B \simeq RTT_b$.  This brings us to Equation \ref{eq:relationship}.

\begin{align} \label{eq:relationship}
B + \bcancel{RTT_B} & < a + RTT_a + b + \bcancel{RTT_b} \nonumber \\
B & < a + RTT_a + b \nonumber \\
B & < a + b + RTT_a
\end{align}

Thus, to determine if Pattern 1.1 is actually better, we need to find values for $B$ (Pattern 1.1 duration for datetime negotiation), $a$ (time to respond to the initial HEAD request in Pattern 2.1), $b$ (Pattern 2.1 duration for datetime negotiation), and $RTT_a$ (the round trip time for the HEAD request during the first step in Pattern 2.1).

\subsubsection{Caching Concerns}

After review of the Wikimedia architecture, it also became apparent that caching was an important aspect of our design and architecture plans.  Because the initial architecture implemented Pattern 2.1 and 302 responses are not supposed to be cached \cite{rfc2616}, caching was not of much concern.  Now that we have decided to pursue Pattern 1.1, it becomes even more important.

Experiments with Varnish (the caching server used by Wikimedia \cite{wikimedia-arch}) indicate that the \emph{Vary} header correctly indicates what representations of the resource are to be cached.  If the URI-R contains a \emph{Vary} header with the value \emph{Accept-Datetime}, this indicates to Varnish that it should cache each URI-R representation in response to an \emph{Accept-Datetime} in the request for that URI-R.  Other values of the \emph{Vary} header have a finite number of values, but \emph{Accept-Datetime} can have a near-infinite number of values (i.e., all datetimes in the past), making caching near useless for Pattern 1.1.

Those visitors of a URI-R that do not use \emph{Accept-Datetime} in the request header will be able to reap the benefits of caching readily.  Memento users of system using Pattern 1.1 will scarcely reap this benefit, because Memento clients send an initial \emph{Accept-Datetime} with every initial request.

Caching is important to our duration equations because a good caching server returns a cached URI-R in a matter of milliseconds, meaning our value of $a$ in Equation \ref{eq:relationship} is incredibly small, on the order of 0.1 seconds on average from our test server.

\subsubsection{Pattern 1.1 vs. Pattern 2.1 URI-G Performance}

The next step was to get a good set of values for $b$, URI-G performance for Pattern 2.1, and $B$, URI-G performance for Pattern 1.1.

\begin{table}
\small
\begin{center}
	\begin{tabular}{| l | l |}
	\hline
	\textbf{CPU Number} & 2 \\
	\hline
	\textbf{CPU Clock Speed} & 2.4 GHz \\
	\hline
	\textbf{CPU Type} & Intel Xeon E7330 \\
	\hline
	\textbf{RAM} & 2 GB \\
	\hline
	\textbf{Operating System} & Red Hat \\
	& Enterprise Linux 6.5 \\
	\hline
	\textbf{Apache HTTP Server Version} & 2.2.15 \\
	\hline
	\textbf{PHP Version} & 5.3.3 \\
	\hline
	\end{tabular}
\end{center}
\caption{Specifications of the Test Machine Used to Compare Pattern 1.1 vs. Pattern 2.1 URI-G Performance}
\label{table:specs}
\end{table}

To get a good range of values, we conducted testing using the benchmarking tool Siege \cite{siege} on our demonstration wiki.  The test machine was a virtual machine with the specifications listed in Table \ref{table:specs}.  The test machine consists of two installs of MediaWiki containing the Memento MediaWiki Extension:  one utilizing Pattern 2.1 and the second implemented using Pattern 1.1.  The data used in the test wikis came from \emph{A Wiki of Ice and Fire}, consisting on many articles about the popular \emph{A Song of Ice and Fire} book series.

Both TimeGate implementations use the same \texttt{TimeNegotiator} class, as shown in the architecture from Figure \ref{fig:version2-classHierarchy}.  They only differ in where this class is called.  The Pattern 1.1 implementation uses the \texttt{ArticleViewHeader} hook \cite{articleviewheader} to instantiate this class and perform datetime negotiation.  The Pattern 2.1 implementation utilizes a MediaWiki \texttt{SpecialPage} \cite{mediawiki-special} at a separate URI to instantiate this class and perform datetime negotiation.

Tests were performed against \emph{localhost} to avoid the benefits of using the installed Varnish caching server.  By doing this, we see the true processing times from MediaWiki for TimeGate response generation.  Also, caching was disabled in MediaWiki to avoid skewing the results.

Siege was run against 6304 different articles in the demonstration wiki.  The date of Mon, 30 Jun 2011 00:00:00 GMT was used for datetime negotiation.  This date corresponds to the release of the book \emph{A Dance With Dragons} which came out after the wiki had an established base of users.  A flurry of activity should occur around and after that date.  All previous books in the \emph{A Song of Ice and Fire} series were released prior to the wiki's creation.

\begin{lstlisting}[frame=single,basicstyle=\scriptsize\ttfamily,captionpos=b,breaklines=true,caption={Example of Siege output},label={lst:siege}, float=*]
HTTP/1.1 302   0.60 secs:       0 bytes ==> GET  /demo-special/index.php/Special:TimeGate/Daenerys
HTTP/1.1 200   3.10 secs:   95662 bytes ==> GET  /demo-special/index.php?title=Daenerys&oldid=27870
HTTP/1.1 302   3.41 secs:       0 bytes ==> GET  /demo/index.php/Daenerys
HTTP/1.1 200   1.86 secs:   94558 bytes ==> GET  /demo/index.php?title=Daenerys&oldid=27870
\end{lstlisting}

Listing \ref{lst:siege} gives an example of the output from Siege.  This output was further processed using a Python script which extracted all of the 302 responses, which correspond to those instances of datetime negotiation (the 200 responses are just Siege dutifully following the 302 redirect).  The URI then indicates which edition of the Memento MediaWiki Extension is installed, differing only in their TimeGate implementation.  URIs beginning with /demo-special use Pattern 2.1.  URIs beginning with /demo use Pattern 1.1.  From these lines we can compare the amount of time it takes to perform datetime negotiation using each design option.  The source code of this script is in Listing \ref{lst:timegateperfpy} in the appendix of this paper.

\begin{figure}[h!]
\centering
	\includegraphics[width=0.5\textwidth]{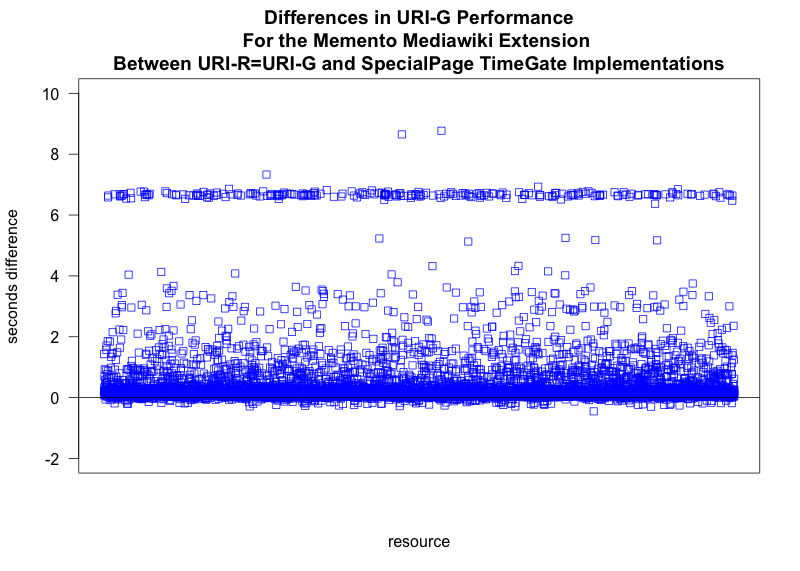}
\caption{Differences in URI-G performance between Pattern 1.1 and 2.1}
\label{fig:URI-G-perf-diff}
\end{figure}

Figure \ref{fig:URI-G-perf-diff} shows the results of this analysis.  The plot shows the difference between the Pattern 1.1 and Pattern 2.1 processing times.  Seeing as most values are above 0, it appears that there is a marked benefit to using Pattern 2.1.  The string of values around 7 seconds difference are all Wiki redirect pages, leading one to infer that redirection is especially expensive with Pattern 1.1.
\begin{figure}[h!]
\centering
	\includegraphics[width=0.5\textwidth]{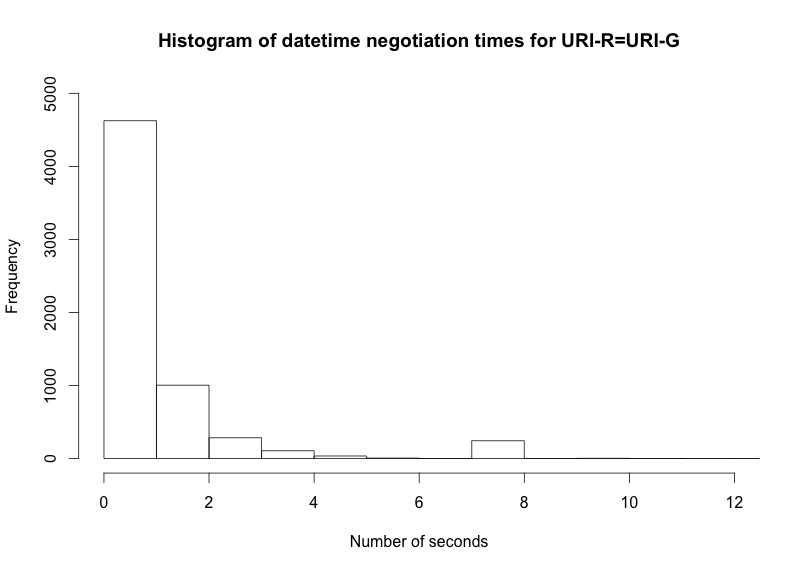}
\caption{Histogram showing Pattern 1.1 values}
\label{fig:pattern1.1-hist}
\end{figure}

Figure \ref{fig:pattern1.1-hist} contains a histogram with 12 buckets containing the range of processing time values for Pattern 1.1.

\begin{figure}[h!]
\centering
	\includegraphics[width=0.5\textwidth]{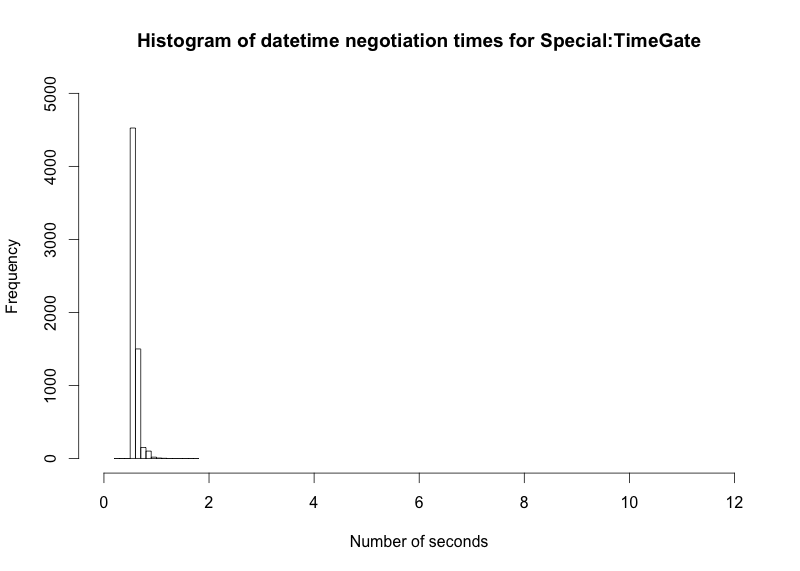}
\caption{Histogram showing Pattern 2.1 values}
\label{fig:pattern2.1-hist}
\end{figure}

Figure \ref{fig:pattern2.1-hist} contains another histogram with 12 buckets for comparison, showing the range of processing time values for Pattern 2.1.
\begin{table}
\small
\begin{center}
	\begin{tabular}{| c | c | c |}
	\hline
	& \textbf{Pattern 1.1} & \textbf{Pattern 2.1} \\
	\hline
	\textbf{Min} & 0.56 & 0.22 \\
	\hline	
	\textbf{Max} & 12.06 & 1.75 \\
	\hline
	\textbf{Mean} & 1.24 & 0.6 \\
	\hline
	\textbf{Median} & 0.77 & 0.59 \\
	\hline
	\end{tabular}
\end{center}
\caption{Statistics on Pattern 1.1 vs. Pattern 2.1 TimeGate testing results}
\label{table:stats}
\end{table}

Table \ref{table:stats} shows the statistics from the testing.  We now have values for $b$ and $B$, so $0.22 \le b \le 1.75$ and $0.56 \le B \le 12.06$ for Equation \ref{eq:relationship}.  Of course, the processing time varies based on page size, number of revisions, and other factors.

The high side of the range of values shown for Pattern 1.1 from Table \ref{table:stats} and Figure \ref{fig:pattern1.1-hist} exceed those acceptable to the MediaWiki performance guidelines \cite{performance-guidelines}.  This also leads one to infer that the cost of using Pattern 1.1 may not be acceptable to the Wikimedia team.

\subsubsection{Round Trip Time}

Our final missing term from Equation \ref{eq:relationship} is $RTT_a$.  RTT is a combination of \emph{transmission delay} ($d_t$),  \emph{propagation delay} ($d_p$), queuing delay, and processing delay \cite{networking-book}.  For the purposes of this paper, we are ignoring queuing delay and processing delay, as those are dependent of the router infrastructure of the Internet and are typically negligible, thus we are reduced to  Equation \ref{eq:rtt}.
\begin{align} \label{eq:rtt}
RTT = d_t + d_p
\end{align}
And transmission delay is a function of the number of bits ($N$) divided by the rate of transmission ($R$) \cite{networking-book}, shown in Equation \ref{eq:dt}.
\begin{align} \label{eq:dt}
d_t = \frac{N}{R}
\end{align}
Listing \ref{lst:example-request-RTTa} shows an example Pattern 2.1 \texttt{HEAD} request.  Considering cookies and other additional data, the average initial Pattern 2.1 \texttt{HEAD} request consists of the 700 Byte HTTP request + a 20 Byte TCP header \cite{tcpip-illustrated1} + a 20 Byte IP header \cite{tcpip-illustrated1}.  This gives a total payload of 740 Bytes or 5920 bits.  Thus our request transmission delay is $d_{trq} = 5920\ b / R$.

\begin{lstlisting}[frame=single,basicstyle=\scriptsize\ttfamily,captionpos=b, breaklines=true, caption={Example HTTP Request for $RTT_a$}, label={lst:example-request-RTTa}, float=*]
HEAD /demo/index.php/Daenerys_Targaryen HTTP/1.1
Host: ws-dl-05.cs.odu.edu
Accept: */*
Accept-Encoding: gzip,deflate,sdch
Accept-Language: en-US,en;q=0.8
Cookie: __utma=99999999.9999999999.9999999999.9999999999.9999999999. 99;
__utmz=99999999.9999999999.9.9.utmcsr=example.com|utmccn=(referral)|utmcmd=referral|utmcct=/; __utma=999999999.9999999999.9999999999.9999999999.9999999999. 9; __utmz=999999999.9999999999.9.9.utmcsr=example|utmccn=(organic)|utmcmd=organic|utmctr=(not%20provided); __atuvc=99%7C99%2C99%7C99%2C9%7C99%2C0%7C99%2C99%7C99
User-Agent: Mozilla/5.0 (Macintosh; Intel Mac OS X 10_9_2) AppleWebKit/537.36 (KHTML, like Gecko) Chrome/34.0.1847.131 Safari/537.36
\end{lstlisting}

\begin{lstlisting}[frame=single,basicstyle=\scriptsize\ttfamily,captionpos=b, breaklines=true, caption={Example HTTP Response for $RTT_a$}, label={lst:example-response-RTTa}, float=*]
HTTP/1.1 200 OK
Age: 0
Cache-Control: s-maxage=18000, must-revalidate, max-age=0
Connection: keep-alive
Content-language: en
Content-Type: text/html; charset=UTF-8
Date: Tue, 06 May 2014 02:57:35 GMT
Last-Modified: Sat, 22 Mar 2014 02:47:30 GMT
Link: <http://ws-dl-05.cs.odu.edu/demo/index.php/Daenerys_Targaryen>; rel="original latest-version",<http://ws-dl-05.cs.odu.edu/demo/index.php/Special:TimeGate/Daenerys_Targaryen>; rel="timegate",<http://ws-dl-05.cs.odu.edu/demo/index.php/Special:TimeMap/Daenerys_Targaryen>; rel="timemap"; type="application/link-format"
Server: Apache/2.2.15 (Red Hat)
Vary: Accept-Encoding,Accept-Datetime,Cookie
Via: 1.1 varnish
X-Content-Type-Options: nosniff
X-Powered-By: PHP/5.3.3
X-Varnish: 2138031585
\end{lstlisting}

Listing \ref{lst:example-response-RTTa} shows an example Pattern 2.1 200 status code reply. Considering variability within the \texttt{Link} header relation entries, the average initial Pattern 2.1 response consists of a 700 Byte HTTP response + a 20 Byte TCP header + a 20 Byte IP header.  This gives a total payload of 740 Bytes or 5920 bits.  Thus our response transmission delay $d_{trs} = 5920\ b / R$.

Seeing as both share the same denominator, our total transmission delay $d_t = d_{trq} + d_{trs} = 5920\ b / R + 5920\ b / R = 11840\ b / R$.

Assuming an average-to-worst case of 1G wireless telephony (28,800 bps), the end user would experience a transmission delay of $d_t = 11840\ b / 28800\ bps = 0.41\ s $.  Combining this with our average case for both TimeGate patterns from the previous section, $b = 0.6\ s$ and $B = 1.24\ s$, and using $a = 0.1$ from the caching results, we get Equation \ref{eq:allbutdp}.
\begin{align} \label{eq:allbutdp}
B & < RTT_a + a + b \textrm{ From (\ref{eq:relationship})} \nonumber \\
B & < d_p + d_t + a + b \textrm{ From (\ref{eq:rtt})} \nonumber  \\
1.24\ s & < d_p + d_t + 0.1\ s + 0.6\ s \nonumber \\
1.24\ s & < d_p + 0.41\ s + 0.1\ s + 0.6\ s \nonumber \\
1.24 & < d_p + 1.11\ s
\end{align}
So, an end user with 1G wireless telephony would need to experience an additional $0.13$ s of propagation delay in order for Pattern 1.1 to be comparable to Pattern 2.1.

Propagation delay is a function of distance and propagation speed, as shown in Equation \ref{eq:dp}.
\begin{align} \label{eq:dp}
d_p = \frac{d}{s_p}
\end{align}
Seeing as wireless telephony travels at the speed of light, the distance one would need to transmit a signal to make Pattern 1.1 viable becomes $80944\ km = 50296.3\ miles$ as shown in Equation \ref{eq:ignoredp}.
\begin{align}\label{eq:ignoredp}
0.13\ s = \frac{d}{299792458\ m/s} \nonumber \\
(0.13\ s)(299792458\ m/s) = d \nonumber \\
d = 38973019.54\ m = 38973\ km = 24216.7\ miles
\end{align}
This is almost the circumference of the Earth \cite{seeds}.  Even if we used copper wire (which has a worse propagation delay) rather than radio waves, the order of magnitude is the same.  Considering the amount of redundancy on the Internet, the probability of hitting this distance is quite low, meaning that propagation delay will likely be so small that we will ignore it for the rest of this discussion.

That brings us back to transmission delay.  At what transmission delay, and essentially what bandwidth, does Pattern 1.1 win out over Pattern 2.1 using our average values for $b$ and $B$?
\begin{align} \label{eq:winout}
B & < d_t + a + b \textrm{ From (\ref{eq:relationship}) and (\ref{eq:rtt}), removing $d_p$} \nonumber \\
1.24\ s & < d_t + 0.1\ s + 0.6\ s \nonumber \\
1.24\ s & < d_t + 0.7\ s \nonumber \\
0.54\ s & < d_t \nonumber \\
d_t & = \frac{N}{R} \textrm{ From (\ref{eq:dt})} \nonumber \\
0.54\ s & = \frac{11840\ b}{R} \nonumber \\
(0.54\ s)(R) & = 11840\ b \nonumber \\
R & = \frac{11840\ b}{0.54\ s} = 21926\ bps
\end{align}
Thus, the bandwidth for which Pattern 1.1 would begin to be useful would be anything at the speed less than 1G telephony, but would become produce increasingly poorer performance for bandwidths higher than that.

\subsubsection{TimeGate Design Conclusion}

From the data gathered and the experiments run, used in Equations \ref{eq:relationship}, \ref{eq:rtt}, and \ref{eq:dt}, Pattern 1.1 takes too much processing time to be viable, in spite of the saved $RTT$.  It comes down to the values of $b$ (processing time for Pattern 2.1) vs. $B$ (processing time for Pattern 1.1), and $B$ is greater in many cases.

Why the big difference?  It turns out that the \texttt{ArticleViewHeader} hook used in the Pattern 1.1 implementation runs after MediaWiki has loaded all of the page data.  The Pattern 2.1 implementation extends a \texttt{SpecialPage}, which has loaded nothing, and can start processing immediately.

Why not use a hook that is run before all of the page data is loaded?  We need a hook that provides MediaWiki's \texttt{WebRequest} object for processing the \emph{Accept-Datetime} request header.  It also needs to provide MediaWiki's \texttt{WebResponse} object for producing the 302 response.  Hooks earlier in the processing chain do not appear to provide this capability.  We prototyped an implementation using the \texttt{BeforeInitialize} hook \cite{beforeinitialize} and it did not preserve the needed response headers, nor did it perform better.  Attempts to find earlier hooks by asking the MediaWiki development team have met with no success \cite{wikitech-question}.

If a MediaWiki hook were available that gave the same performance for Pattern 1.1 as for Pattern 2.1 then transmission delay would no longer matter, and Pattern 1.1 would clearly be the best choice, as we see from Equation \ref{eq:gopat11}, because transmission delay would always be greater.
\begin{align} \label{eq:gopat11}
B & < d_t + a + b \textrm{ From (\ref{eq:relationship}) and (\ref{eq:rtt}), removing $d_p$} \nonumber \\
b & < d_t + 0.1\ s + b \textrm{ Replacing $B$ with mean of $b$} \nonumber \\
b - b & < d_t + 0.1\ s + b - b \nonumber \\
0 & < d_t + 0.1\ s
\end{align}
Of course, the processing time is not the only issue here; the use of Pattern 1.1 would make caching useless for Memento users of URI-Rs, considering Memento clients send an \emph{Accept-Datetime} with each request, and there are a near infinite number of values for \emph{Accept-Datetime}.



\begin{table*}[t]
\small
\begin{center}
	\begin{tabular}{| l | l | l |}
	\hline
	\textbf{Condition} & \textbf{Status Code} & \textbf{Reasoning} \\
	\hline
	Special:TimeGate was requested without any article name & 200 & This way administrators and visitors can learn \\
	& & how Special:TimeGate is used \\
	\hline
	Datetime Negotiation Was Successful & 302 & As detailed in RFC 7089 Pattern 2.1 \\
	\hline
	Supplied string in Accept-Datetime is not formatted correctly, & 400 & As detailed in RFC 7089 section 4.5.3 \\
	or contains data that is incorrect (e.g., Feb. 30) & & \\
	\hline
	Datetime negotiation is not available for the given namespace & 403 & The extension is refusing to fulfill the request \\
	\hline
	Given article name does not exist & 404 & MediaWiki cannot find anything matching this article \\
	\hline
	\end{tabular}
\end{center}
\caption{HTTP Status Codes for a Memento MediaWiki Extension TimeGate, if \texttt{\$wgMementoErrorPageType} is set to \texttt{'traditional'}}
\label{tab:statusTimeGate}
\end{table*}

\begin{table*}[t]
\small
\begin{center}
	\begin{tabular}{| l | l | l |}
	\hline
	\textbf{Condition} & \textbf{Status Code} & \textbf{Reasoning} \\
	\hline
 	Special:TimeMap was requested with a valid article name & 200 & This is the successful known-good state \\
	\hline
	Supplied pivot string in URI is not formatted correctly, & 400 & As detailed in RFC 7089 section 4.5.3 \\
	or contains data that is incorrect & & \\
	& & (e.g., 20140230000000, which is Feb. 30) \\
	\hline
	TimeMaps are not available for the given namespace & 403 & The extension is refusing to fulfill the request \\
	\hline
	Given article name does not exist & 404 & MediaWiki cannot find anything matching this article \\
	\hline
	\end{tabular}
\end{center}
\caption{HTTP Status Codes for a Memento MediaWiki Extension TimeMap, if \texttt{\$wgMementoErrorPageType} is set to \texttt{'traditional'}}
\label{tab:statusTimeMap}
\end{table*}

\subsection{Installation Options}

The Memento MediaWiki Extension is installed by uncompressing the source into MediaWiki's \texttt{extensions} directory and adding the code from Listing \ref{lst:install-lines} to the \texttt{LocalSettings.php} file.
\begin{lstlisting}[frame=single,basicstyle=\scriptsize\ttfamily,captionpos=b,breaklines=true,caption={Lines to add to \texttt{LocalSettings.php} to enable the Memento MediaWiki Extension},label={lst:install-lines}, float=*]
require_once( "$IP/extensions/Memento/Memento.php" );
$wgArticlePath = "$wgScriptPath/index.php/$1";
$wgUsePathInfo = true;
\end{lstlisting}

While this will enable the Memento MediaWiki Extension with the default options, the install can be configured using the global variables shown in Table \ref{tab:configoptions}.

\begin{table*}[t]
\small
\begin{center}
	\begin{tabular}{| l | l | l | l |}
	\hline
	\textbf{Configuration} & \textbf{Description} & \textbf{Possible} & \textbf{Default} \\
	\textbf{Option} & & \textbf{Values} & \textbf{Value} \\
	\hline
	\texttt{\$wgMementoTimemapNumberOfMementos} & allows the admin to alter the number & any integer & \texttt{500} \\
	& of Mementos in a TimeMap  & value $i > 0$ & \\
	\hline
	\texttt{\$wgMementoErrorPageType} & allows the admin to choose between & string values of  & \texttt{'friendly'} \\
	& 'traditional' (actual 4** and 5** status codes) & \texttt{'friendly'} or & \\
	& and 'friendly' (200 status with error message in body)& \texttt{'traditional'} & \\
	& error message & & \\
	\hline
	\texttt{\$wgMementoTimeNegotiation} & allows the admin to change the & string values of & \texttt{'302'} \\
	& datetime negotiation pattern; &  \texttt{'302'} or & \\
	& 302 corresponds to Pattern 2.1 & \texttt{'200'} & \\
	& 200 corresponds to Pattern 1.2 & & \\
	\hline
	\texttt{\$wgMementoRecommendedRelations} & allows the admin to enable or disable & boolean values of & \texttt{false} \\
	& \emph{recommended} relations as defined by RFC 7089; & \texttt{false} or & \\
	& \texttt{true} enables \emph{all} Memento headers & \texttt{true} & \\
	& \texttt{false} only enables \emph{mandatory} Memento headers & & \\
	\hline
	\texttt{\$wgMementoExcludeNamespaces} & allows the admin to exclude & any integer & all values but \\
	& certain MediaWiki namespaces from & value $i > 0$ & 0 (NS\_MAIN) \\
	& the extension & & \\
	\hline
	\end{tabular}
\end{center}
\caption{Memento MediaWiki Extension Configuration Options and meanings}
\label{tab:configoptions}
\end{table*}

The \texttt{\$wgMementoTimemapNumberOfMementos} setting restricts the number of Mementos returned in a TimeMap.  This setting was implemented in version 1.0 due to concerns by Wikimedia that TimeMap processing time could be considerable for wiki pages containing thousands of revisions, hence thousands of entries in the TimeMap.  The value of $500$ was chosen as a sensible default because it is the same number limit on values returned by calls to the MediaWiki API.

The \texttt{\$wgMementoErrorPageType} setting allows an administrator to choose between ``friendly'' and ``traditional'' error pages.  This was implemented to conform to MediaWiki's design and coding standards.  In this context, an error page is what results from some kind of poor server condition or as a result of bad input.  Typically HTTP servers return 4** and 5** status codes in response to these conditions.  A ``traditional'' error page preserves these status codes (e.g., returning a 404 status code for a page not found), and the extension acts as shown in Tables \ref{tab:statusTimeGate} and \ref{tab:statusTimeMap}.  A ``friendly'' error page returns a 200 status code containing the error message inside the page body.

The \texttt{\$wgMementoTimeNegotiation} setting allows the administrator to select the datetime negotiation pattern to use.  Based on the results of the experiments in the previous section, the only 302-style negotiation pattern is Pattern 2.1 shown in Figure \ref{fig:salient-pattern2.1}, hence a string value of \texttt{302} will enable it.  Alternatively, one can select the 200-style Pattern 1.2 shown in Figure \ref{fig:salient-pattern1.2} by using a string value of \texttt{200}.  Pattern 1.2 was included because it was assumed that the Wikimedia developers would prefer it, instead, in retrospect, they suggested that the 302-style pattern would be a better default \cite{extension-advice-1, extension-advice-2}.

The \texttt{\$wgMementoExcludeNamespaces} setting allows the administrator to disable datetime negotiation for specific wiki namespaces.  The Wikimedia team suggested that this setting contain namespaces that are not \emph{content namespaces}.  Content namespaces are those reserved for actual wiki pages.  By default, wiki pages reside in the namespace NS\_MAIN.  The default value for this setting enables datetime negotiation for NS\_MAIN (i.e., value of $0$), but disables it for all other namespaces, such as Talk pages or User pages.

The \texttt{\$wgMementoRecommendedRelations} setting allows the administrator to enable or disable the \emph{recommended} relations typically included in the \texttt{Link} header for Memento.  If this setting is turned on, the relations \emph{first memento} and \emph{last memento} are added to URI-M and URI-R headers and TimeMap output.  For the rest of this tech report, the term \emph{all headers installation} refers to the state where this setting is set to \texttt{true} and all of these additional relations are generated and returned as part of the server response header.

\section{Performance Impact on \newline MediaWiki Installations}

Once we completed initial development on the Memento MediaWiki Extension, we turned our focus to its impact on performance.  We used Siege again, as in the TimeGate design experiment.  The same machine as shown in Table \ref{table:specs} was used to run these performance tests, and the same demonstration wiki provided the test data.

As URI-Gs were tested during the TimeGate design experiment, we focused our attention on the other Memento resource types.

The Python code in Listing \ref{lst:urirurimperfpy} at the end of this paper was used to process the URI-R and URI-M test results.

\subsection{URI-R Performance}

First, we look at the results for URI-Rs.  These are the base wiki article pages.  All the Memento MediaWiki Extension does is add Memento headers to these pages for a Memento client's benefit, informing the client of the URI for the TimeGate and TimeMap, and, in the case where all headers are enabled, first and last mementos.

\begin{figure*}
\centering

\begin{subfigure}{1.0\textwidth}
\centering
	\includegraphics[width=0.5\textwidth]{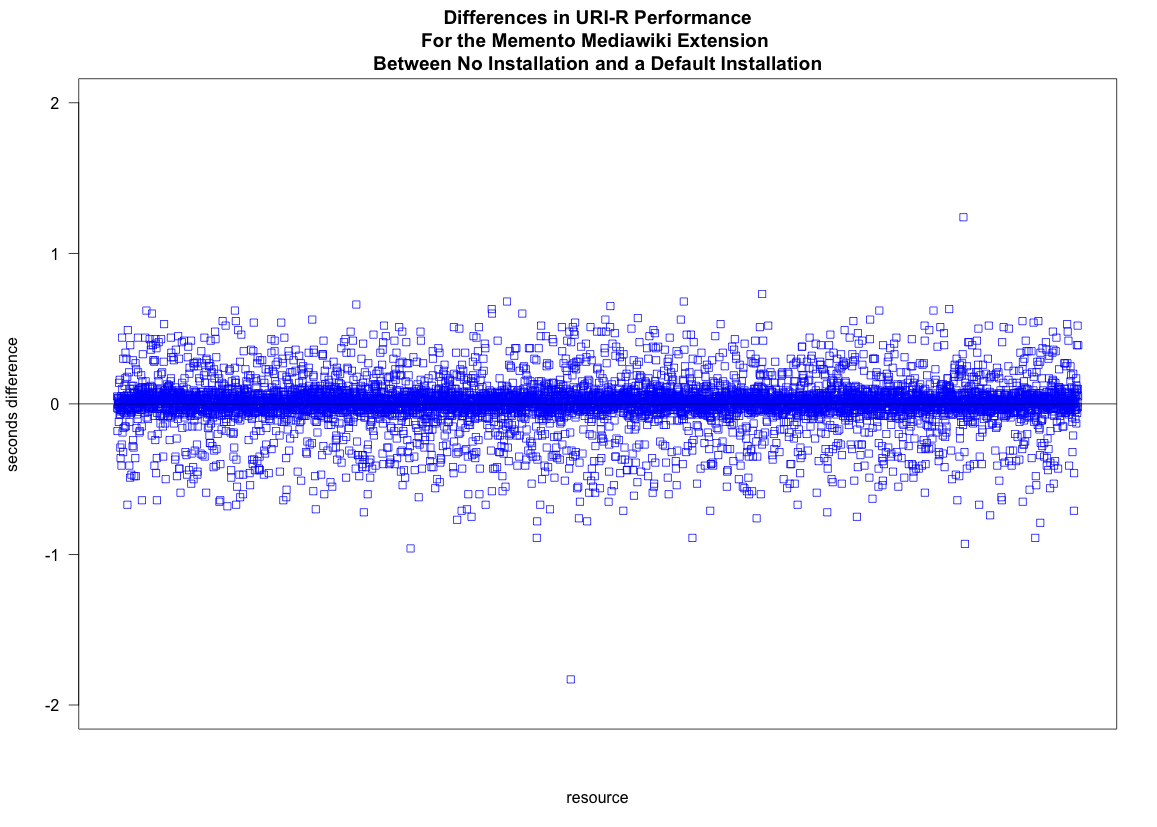}
\caption{Plot showing the difference in times for URI-Rs between a Memento MediaWiki Extension installation with only mandatory headers and no install}
\label{fig:urir-default-no}
\end{subfigure}%

\begin{subfigure}{1.0\textwidth}
\centering
	\includegraphics[width=0.5\textwidth]{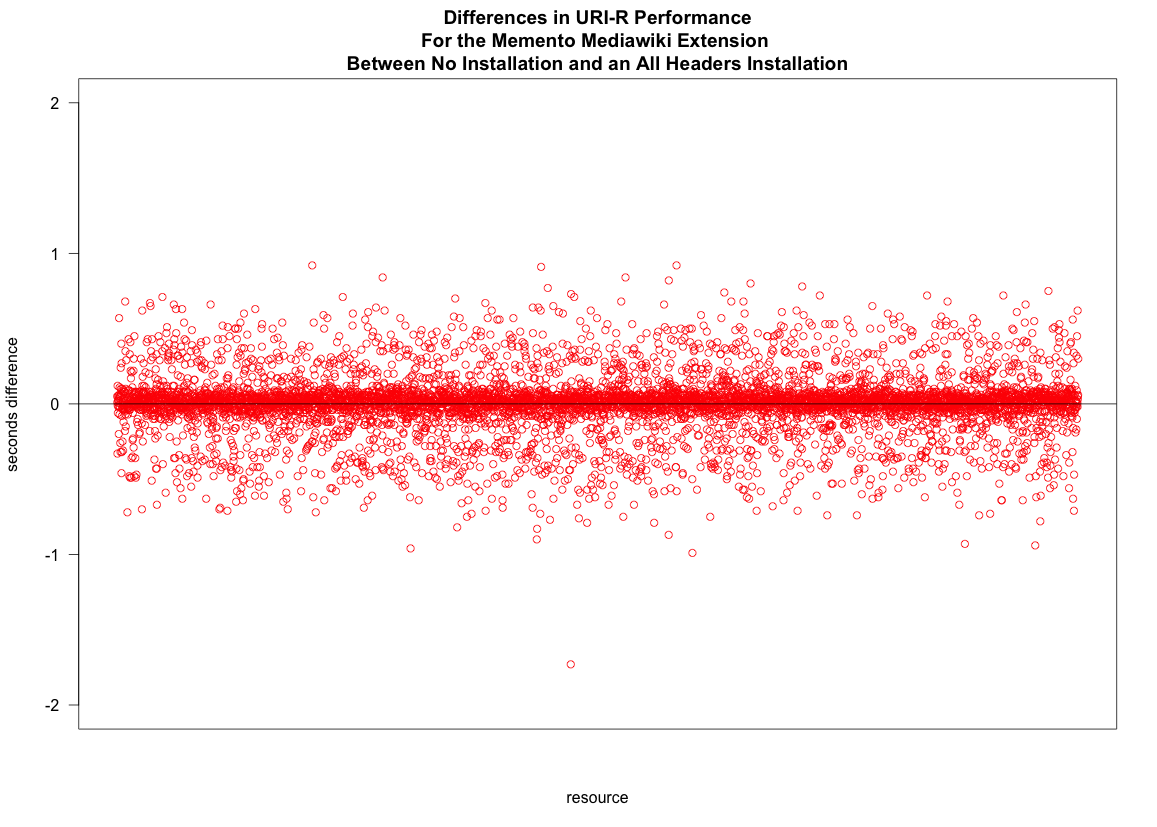}
\caption{Plot showing the difference in times for URI-Rs between a Memento MediaWiki Extension installation with all headers turned on and no install}
\label{fig:urir-all-no}
\end{subfigure}%

\begin{subfigure}{1.0\textwidth}
\centering
	\includegraphics[width=0.5\textwidth]{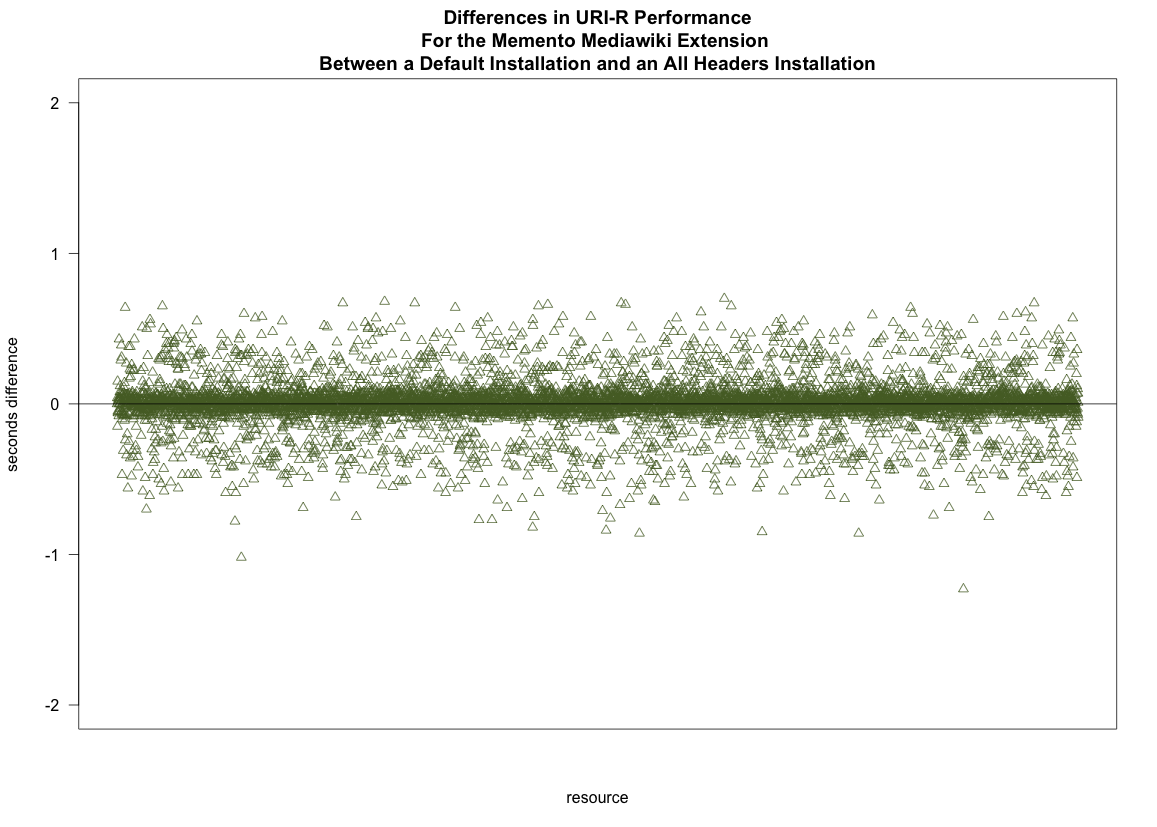}
\caption{Plot showing the difference in times for URI-Rs between a Memento MediaWiki Extension installation with all headers turned on and one with only mandatory headers}
\label{fig:urir-all-default}
\end{subfigure}

\caption{Results of URI-R performance testing}
\label{fig:urir-testing}
\end{figure*}

Figure \ref{fig:urir-default-no} shows the difference in seconds between accessing a wiki page's URI-R with the Memento MediaWiki Extension installed and accessing the same wiki page without the extension loaded.  Each point on the plot is one of 6480 different pages from the test wiki. The plots are evenly arranged around the 0 mark, with most of the points between 0.5 and -0.5.  This means that installing the extension has a negligible impact on performance of URI-Rs.  If the extension seriously impacted performance, then most of the plots should be above the 0 mark.

Figure \ref{fig:urir-all-no} shows the difference in seconds between accessing a wiki page's URI-R with the Memento MediaWiki installed with all headers turned on and accessing the same wiki page without the extension loaded.  Each point on the plot is again, one of 6480 different pages from the test wiki, and again they are evenly arranged around the 0 mark.  This time, it appears most of the points aer between 0.7 and -0.7, but they are still spread rather evenly around 0.  Because most of the points are around the 0 mark, using the extension with all headers enabled still should have a negligible impact on performance.

Figure \ref{fig:urir-all-default} shows different information.  It shows the performance difference between an install with all headers enabled and only one with mandatory headers enabled.  It was hypothesized that enabling the headers would cause performance issues with the system, but as the data shows, the difference is still very small, with data points on either side of the 0 mark.

\subsection{URI-M Performance}

\begin{figure*}
\centering
\begin{subfigure}{1.0\textwidth}
\centering
	\includegraphics[width=0.5\textwidth]{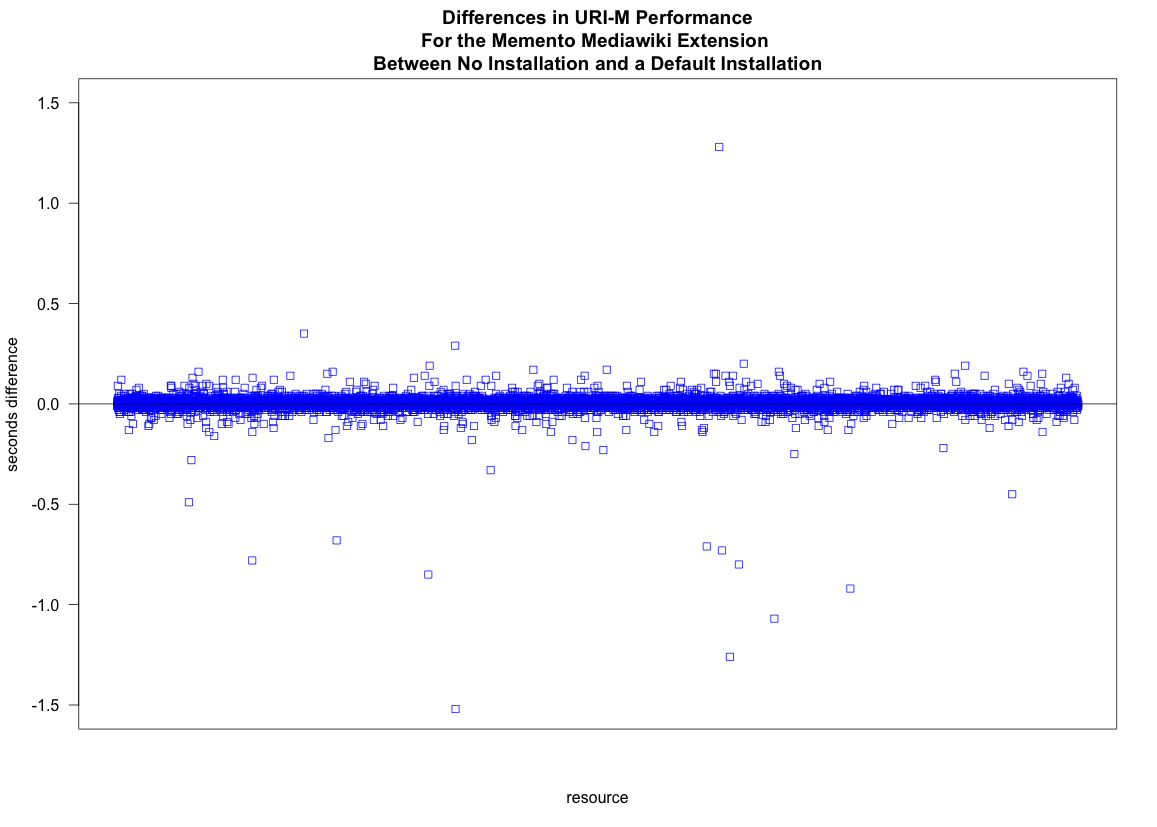}
\caption{Plot showing the difference in times for URI-Ms between a Memento MediaWiki Extension installation with only mandatory headers enabled and no install}
\label{fig:urim-default-no}
\end{subfigure}

\begin{subfigure}{1.0\textwidth}
\centering
	\includegraphics[width=0.5\textwidth]{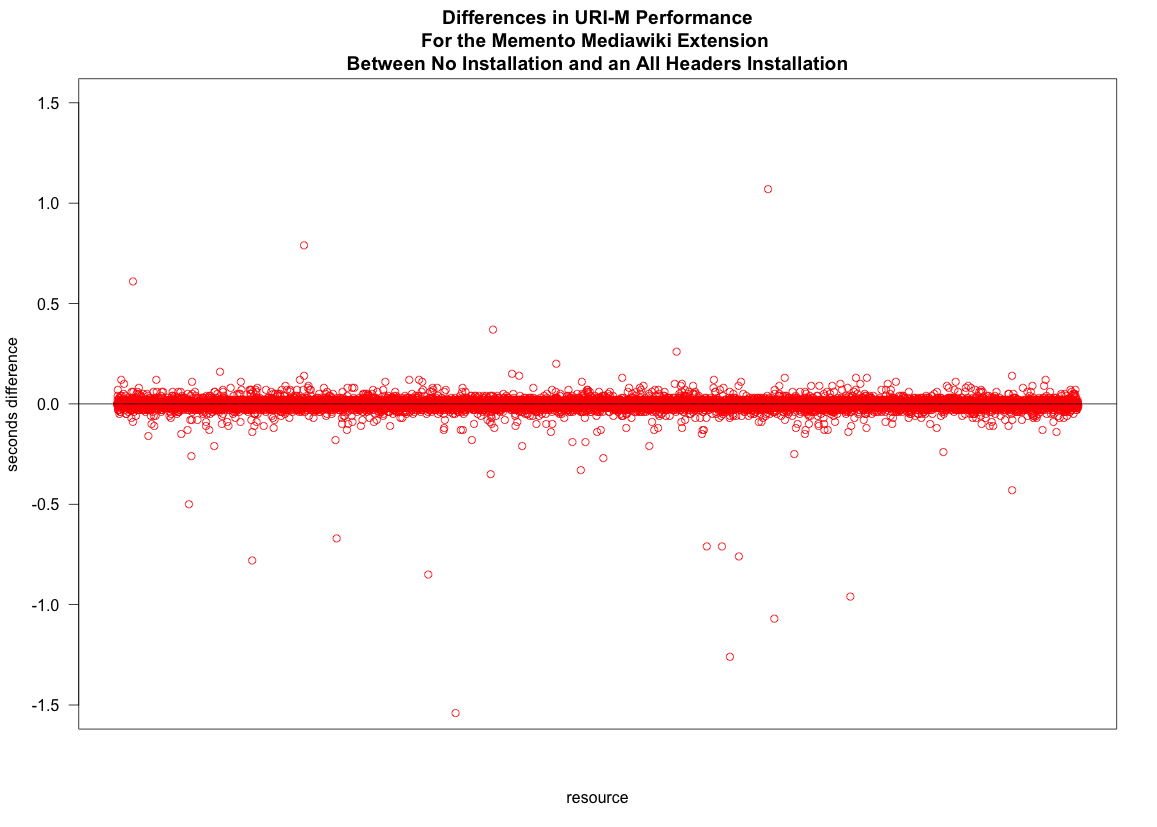}
\caption{Plot showing the difference in times for URI-Ms between a Memento MediaWiki Extension installation with all headers turned on and no install}
\label{fig:urim-all-no}
\end{subfigure}

\begin{subfigure}{1.0\textwidth}
\centering
	\includegraphics[width=0.5\textwidth]{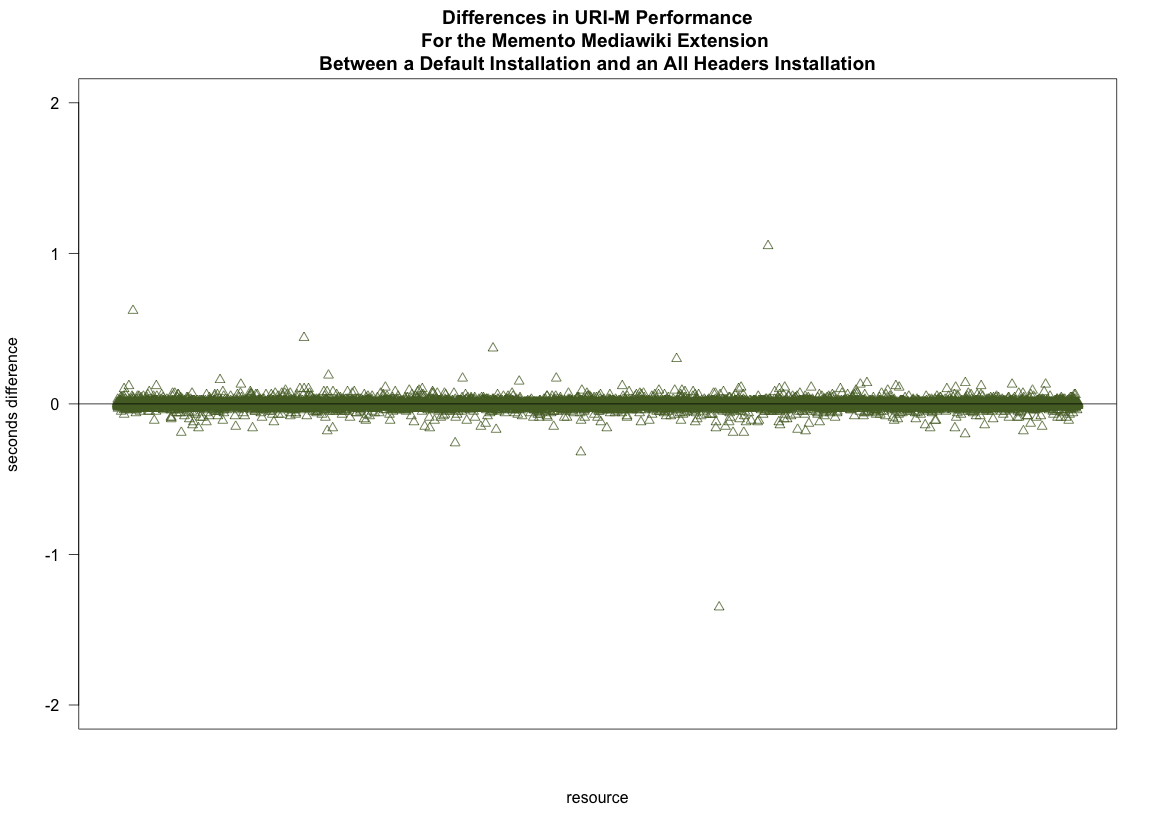}
\caption{Plot showing the difference in times for URI-Ms between a Memento MediaWiki Extension installation with all headers turned on and with only mandatory headers}
\label{fig:urim-all-default}
\end{subfigure}
\caption{Results of URI-M performance testing}
\label{fig:urim-testing}
\end{figure*}

Secondly, we look at the results for URI-Ms, or \emph{oldid pages}.  This is the other Memento resource type that MediaWiki natively implemented already.  Just like with URI-Rs, the Memento MediaWiki Extension adds Memento headers to these pages for a Memento client's benefit, informing the client of the URI for the TimeGate and TimeMap, and, in the case where all headers are enabled, first and last mementos.

Figure \ref{fig:urim-default-no} shows the difference in seconds between accessing a URI-M (or oldid page in MediaWiki parlance) with only mandatory Memento headers enabled and accessing the same page without the extension installed.  Each point on the plot is one of 10257 different oldid pages from the test wiki.  These plots are also arranged around the 0 mark, with most of the points between -0.25 and 0.25.  This means that installing the extension has a negligible impact on URI-Ms.  Again, if the extension seriously impacted performance, then most of the plots should be above the 0 mark.

Figure \ref{fig:urim-all-no} shows the same difference, but with all headers enabled.  Again, we see most points clustered around either side of the 0 mark, indicating a minimal impact to performance for URI-Ms.

Figure \ref{fig:urim-all-default} shows the difference in performance between an all headers installation and one with only mandatory headers.  Again, turning on all of the headers makes a minimal impact to performance versus only using the defaults.  This was unexpected, as we again hypothesized that the calculation time needed to generate these additional headers would have a large impact on performance.

\subsection{URI-T Performance}

The closest thing to a Memento TimeMap (URI-T) in MediaWiki is a history page, but they are not really the same thing.  The audience for history pages are humans, whereas the audience for TimeMaps are machine clients.  Seeing as 80.8\% of requests for TimeMaps come from machine clients \cite{links-ia}, and 95\% of machine clients download TimeMaps exclusively \cite{access-patterns}, there is interest in providing a machine readable format of the history page.  To use a history page, a machine client would need to parse the HTML, performing unnecessary computation in order to get the same data provided much more succinctly by a TimeMap.

Again, we used Siege to download 6252 sample history pages and TimeMaps from our demonstration wiki.  The Python code shown in the Appendix under Listing \ref{lst:uritperfpy} was used to process this data.

\begin{figure}[h!]
\centering
	\includegraphics[width=0.5\textwidth]{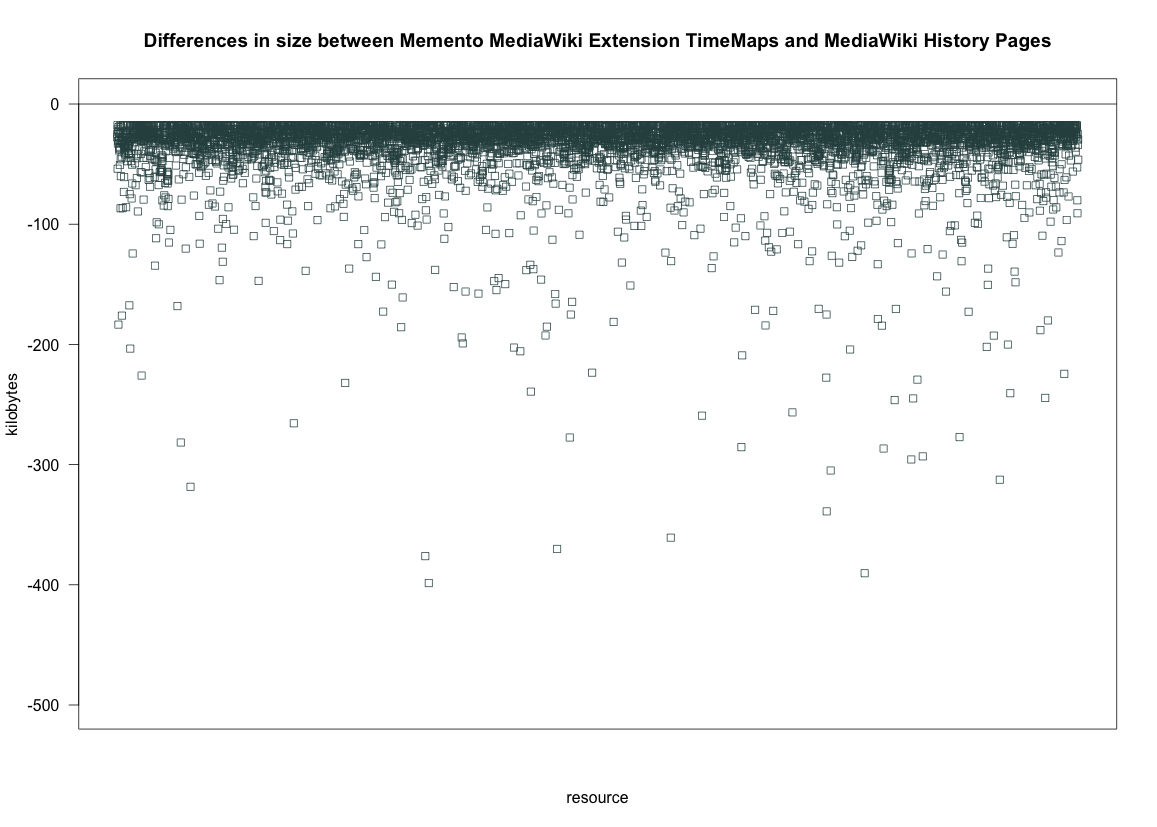}
\caption{Plot showing the difference in size between MediaWiki history pages and TimeMaps for the same article}
\label{fig:urit-bytes}
\end{figure}

Figure \ref{fig:urit-bytes} shows the difference in size between a MediaWiki history page and the corresponding TimeMap for the same article across 6252 sample pages.  The mean in this sample is -34.7 kilobytes.  This means, that if one were to solely rely upon a MediaWiki history page to acquire TimeMap data, they would need to parse through an additional unnecessary 35 kilobytes.  In addition, there would be extra processing time given to stripping out the HTML and generating the TimeMap, which is a waste when a standard format TimeMap exists already.

Of course, one could also use the MediaWiki API to generate the information for TimeMaps, but the API limits one to 500 records \cite{mediawiki-api}, whereas TimeMaps provide paging and allow one to browse beyond this limit.  Additionally, TimeMaps provide URIs, whereas the MediaWiki API provides revision identifiers, which would require one to construct URIs in addition to parsing the API output in order to produce a TimeMap.

\section{Additional Considerations}

\begin{table}
\small
\begin{center}
	\begin{tabular}{| l | l |}
	\hline
	\textbf{MediaWiki Entity} & \textbf{Status of Solution for Memento} \\
	\hline
	Wiki Article & Complete in Extension \\
	\hline
	Template Page & Complete in Extension \\
	\hline
	Embedded Images & Prototyped for next version of Extension \\
	\hline
	Embedded JavaScript & Requires change to MediaWiki source \\
	\hline
	Embedded StyleSheets & Requires change to MediaWiki source \\
	\hline
	\end{tabular}
\end{center}
\caption{Status of full temporal coherence among MediaWiki Entities}
\label{tab:mwentities}
\end{table}

Of course, the Memento MediaWiki Extension works fine for extracting previous versions of pages, as well as the MediaWiki templates that go with them, but we want to achieve true \emph{temporal coherence} \cite{temporal-coherence}.

Web archives process a web page and retrieve the embedded resources at some point thereafter, which creates all kinds of problems when attempting to reconstruct the page to resemble its past revision \cite{ainsworth-nelson-temporaldrift}.  MediaWiki has access to every revision of its embedded resources, therefore true temporal coherence should be achievable.  To realize this, each MediaWiki URI-M must contain all of the correct revisions of those embedded images, JavaScript, and stylesheets that existed at the time the URI-M was saved.  Table \ref{tab:mwentities} shows the status of this work.

As we show below, the temporal coherence of all Mementos served by MediaWiki is potentially a condition called \emph{prima facia violative}, specifically the pattern \emph{Right Newer Last-Modified}.  This means that past revisions of a MediaWiki page contain the current revision of embedded resources.

The following sections highlight the issues of MediaWiki's temporal coherence in more detail.

\subsection{Embedded Images}

One of the problems we seek to address is the issue of embedded images \cite{blog-yesterdays-page-todays-image}.  MediaWiki allows one to store multiple versions of an embedded image under a single page name in the \emph{File} namespace.

\begin{figure}[h!]
\centering
	\includegraphics[width=0.5\textwidth]{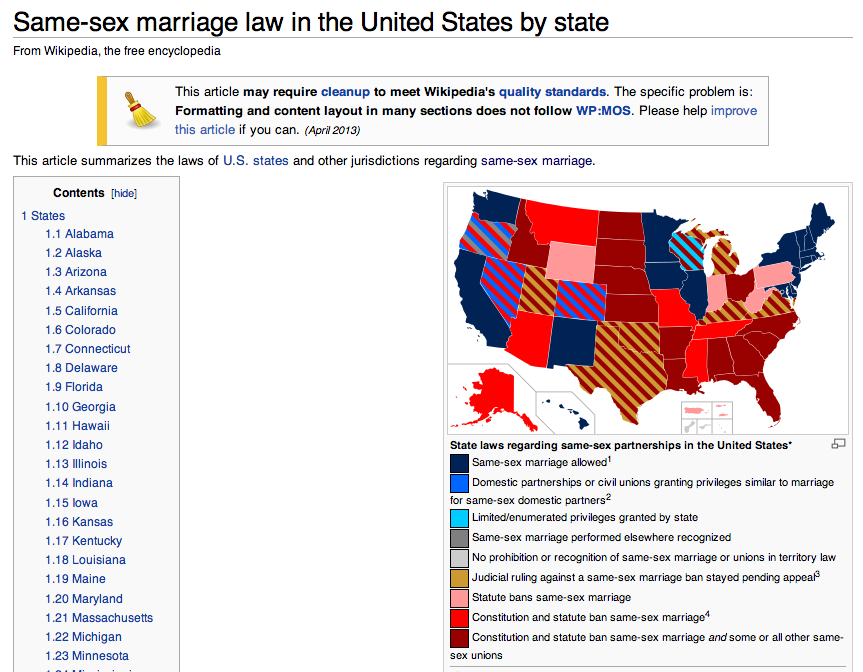}
\caption{Example MediaWiki page with an embedded image that has been changed as the page content changes}
\label{fig:example-wikipedia-current}
\end{figure}

Figure \ref{fig:example-wikipedia-current} shows a screenshot of a Wikipedia page containing a map showing the legal status of Same-sex marriage law in the United States.  The article content is changed as this issue unfolds, and the map is updated also to reflect the article content.

If we access previous revisions of the MediaWiki page now, then it displays the \emph{current} revision of the map, \emph{not the one that goes with that revision of the article}.

\begin{figure}[h!]
\centering
	\includegraphics[width=0.5\textwidth]{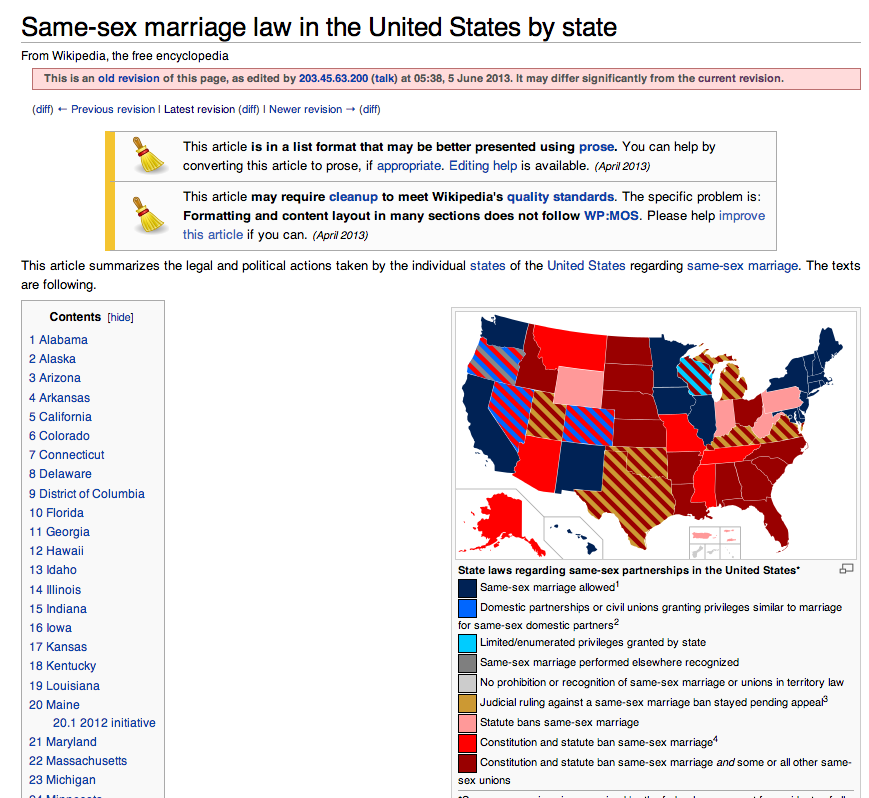}
\caption{July 5, 2013 version of the example MediaWiki page with an embedded image that is changed as the page content changes (note that the map is the same as in Figure \ref{fig:example-wikipedia-current}, which does not match the article text)}
\label{fig:example-wikipedia-old}
\end{figure}

\begin{figure}[h!]
\centering
	\includegraphics[width=0.5\textwidth]{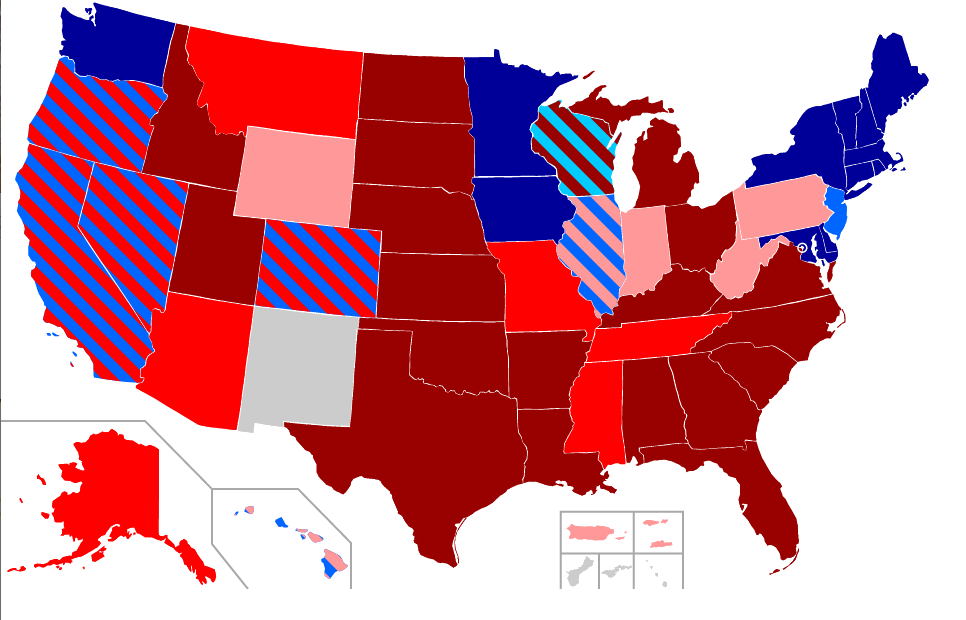}
\caption{July 5, 2013 version of the example MediaWiki page should show this map instead if it is to be consistent with the article content}
\label{fig:correct-map}
\end{figure}

What should be shown is the image shown in Figure \ref{fig:correct-map} because it accurately reflects the content of the July 5, 2013 revision of the article.

\begin{figure}[h!]
\centering
	\includegraphics[width=0.4\textwidth]{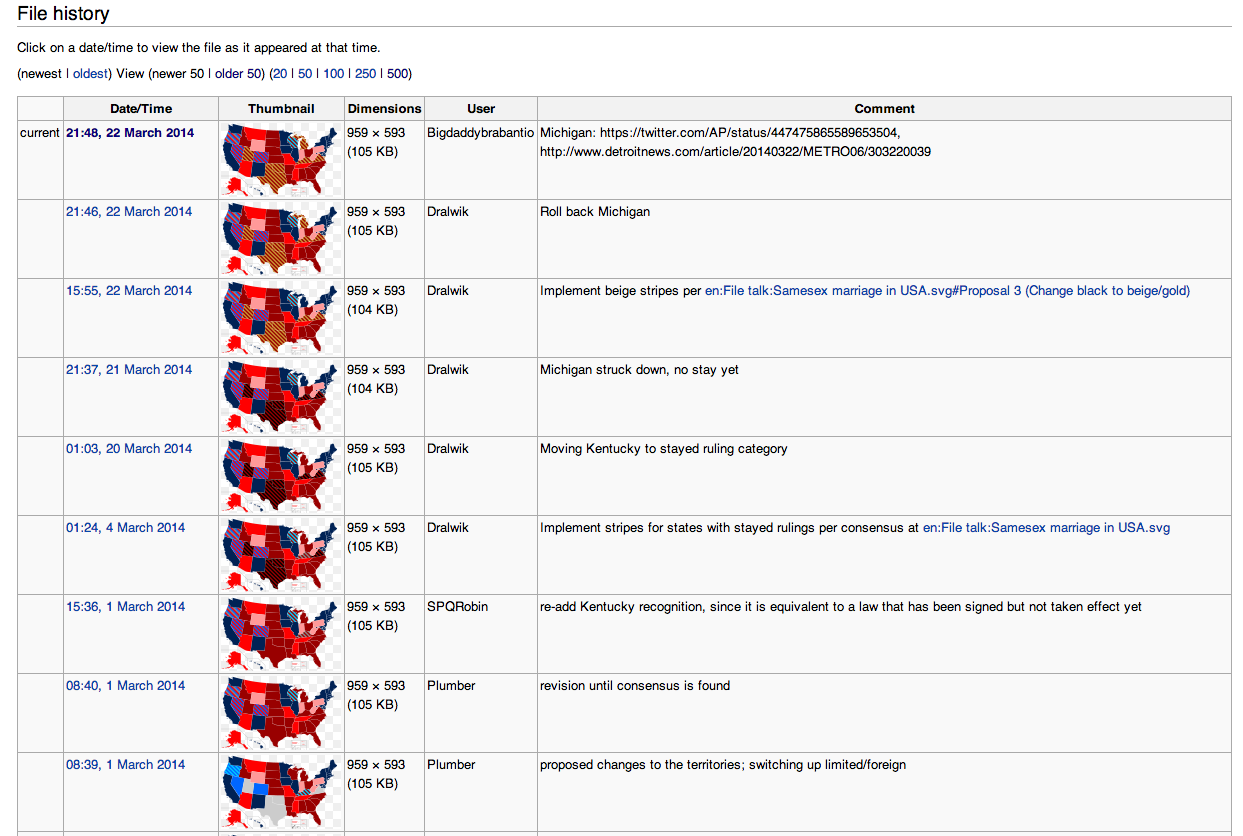}
\caption{MediaWiki Page showing the map's file history}
\label{fig:map-file-history}
\end{figure}

Figure \ref{fig:map-file-history} shows that Wikipedia (and transitively, MediaWiki) has access to all of the previous revisions of the map.  The data is present in the system, but MediaWiki does not present the previous version of the image with the previous version of the page.

MediaWiki provides the \texttt{ImageBeforeProduceHTML} hook,\newline which provides a \texttt{\$file} argument, giving access to the \texttt{LocalFile} object for the embedded image.  It also provides a \texttt{\$time} argument that signifies the \emph{Timestamp of file in 'YYYYMMDDHHIISS' string form, or false for current} \cite{imagebeforeproducehtml}.

We wanted to use the \texttt{\$time} argument, but were perplexed when the hook did not perform as expected, so we examined the source of MediaWiki version 1.22.5. Listing \ref{lst:linker-hookcall} shows the hook being called within the MediaWiki file Linker.php.

Listing \ref{lst:linker-timepass1} shows that the \texttt{\$time} variable that we would set is passed to the \texttt{makeThumbLink2} function, also in the same file.

\begin{lstlisting}[frame=single,basicstyle=\scriptsize\ttfamily,captionpos=b,numbers=left,breaklines=true,caption={Where the ImageBeforeProduceHTML hook is called in Linker.php},label={lst:linker-hookcall},firstnumber=569, float=*, language=PHP]
 if ( ~{\color{red}!wfRunHooks( 'ImageBeforeProduceHTML', array( \&\$dummy, \&\$title,
   \&\$file, \&\$frameParams, \&\$handlerParams, \&\$time, \&\$res ) )~ ) {
   return $res;
  }
\end{lstlisting}

\begin{lstlisting}[frame=single,basicstyle=\scriptsize\ttfamily,captionpos=b,numbers=left,breaklines=true,caption={Where the variable \texttt{\$time} is passed after the hook is called},label={lst:linker-timepass1},firstnumber=639, float=*, language=PHP]
  if ( isset( $fp['thumbnail'] ) || isset( $fp['manualthumb'] ) || isset( $fp['framed'] ) ) {
   # Create a thumbnail. Alignment depends on the writing direction of
   # the page content language (right-aligned for LTR languages,
   # left-aligned for RTL languages)
   #
   # If a thumbnail width has not been provided, it is set
   # to the default user option as specified in Language*.php
   if ( $fp['align'] == '' ) {
    if ( $parser instanceof Parser ) {
     $fp['align'] = $parser->getTargetLanguage()->alignEnd();
    } else {
     # backwards compatibility, remove with makeImageLink2()
     global $wgContLang;
     $fp['align'] = $wgContLang->alignEnd();
    }
   }
   ~{\color{red} return \$prefix . self::makeThumbLink2( \$title, \$file, \$fp, \$hp, \$time, \$query ) . \$postfix;}~
  }
\end{lstlisting}

\begin{lstlisting}[frame=single,basicstyle=\scriptsize\ttfamily,captionpos=b,numbers=left,breaklines=true,caption={Where the variable \texttt{\$time} is used to create a boolean value},label={lst:linker-timebool1},firstnumber=861, float=*, language=PHP]
  if ( !$exists ) {
   ~{\color{red} \$s .= self::makeBrokenImageLinkObj( \$title, \$fp['title'], '', '', '', \$time == true );}~
   $zoomIcon = '';
  } elseif ( !$thumb ) {
\end{lstlisting}

\begin{lstlisting}[frame=single,basicstyle=\scriptsize\ttfamily,captionpos=b,numbers=left,breaklines=true,caption={Where the variable \texttt{\$time} is again used to create a boolean value},label={lst:linker-timebool2},firstnumber=674, float=*, language=PHP]
if ( !$thumb ) {
 ~{\color{red} \$s = self::makeBrokenImageLinkObj( \$title, \$fp['title'], '', '', '', \$time == true );}~
} else {

\end{lstlisting}

But, as shown in Listing \ref{lst:linker-timebool1}, the value of \texttt{\$time} is not really used.  Instead, it is used to create a boolean value before being passed on to \texttt{makeBrokenLinkObj}.

Back inside the \texttt{makeImageLink} function, we see a second use of the \texttt{\$time} value, as shown in Listing \ref{lst:linker-timebool2}, but it is again used to create a boolean argument to the same function as seen in Listing \ref{lst:linker-timebool1}.

Note that its timestamp value of \texttt{\$time} in 'YYYYMMDDHHIISS' string form is never actually used as described.  So, the documentation for the ImageBeforeProduceHTML hook is incorrect on the use of this \texttt{\$time} argument.  In fact, the hook was introduced in MediaWiki version 1.13.0 and this code does not appear to have changed much since that time.  It is possible that the \texttt{\$time} functionality is intended to be implemented in a future version.

Finally, we discovered a possible solution by instead using the \texttt{\$file} object's \texttt{getHistory()} function \cite{getHistory}.  This function returns an array of the \texttt{File} objects representing each revision of an image.  Even better, it takes \texttt{\$start} and \texttt{\$end} arguments, meaning that this function can do the datetime negotiation itself.  Seeing as the \texttt{\$file} argument is passed in by reference to the \texttt{ImageBeforeProduceHTML}, we can reassign the \texttt{File} object to the one in the array with the desired datetime, thus loading the correct image.

Our final solution requires more review, as one needs to purge the MediaWiki cache in order to view the correct revision of the image.  We also need to determine how to retrieve the correct datetime for the URI-M base page that loads the image.  For these reasons, images are not currently supported by the extension, but as noted in Table \ref{tab:mwentities}, this capability has been prototyped for the next version of the Memento MediaWiki Extension.

\subsection{Embedded JavaScript and CSS}

\begin{figure}[h!]
\centering
	\includegraphics[width=0.5\textwidth]{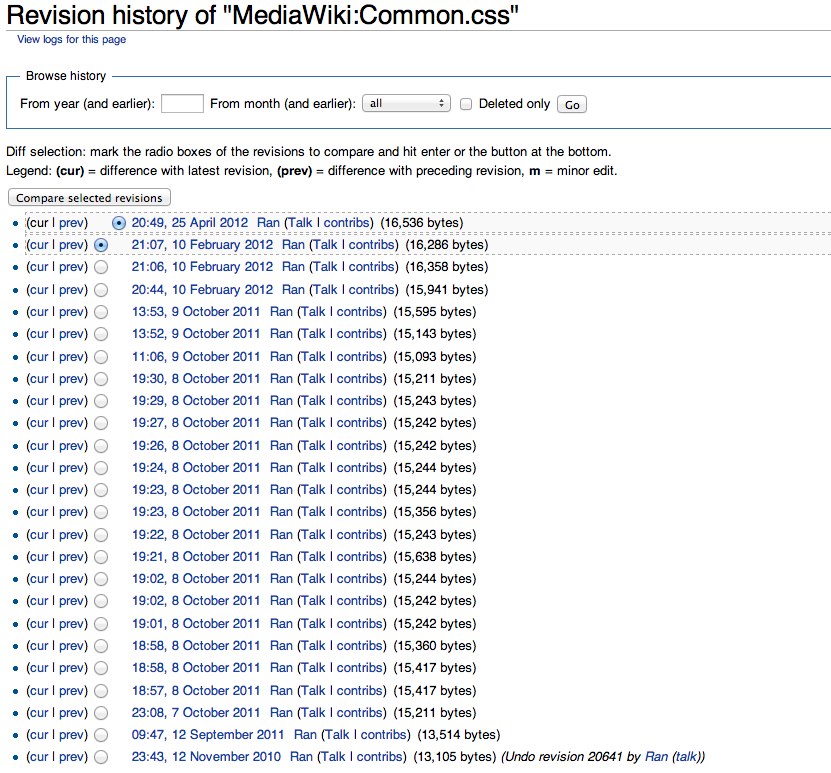}
\caption{Example of CSS history in MediaWiki}
\label{fig:css-history}
\end{figure}

\begin{figure}[h!]
\centering
	\includegraphics[width=0.5\textwidth]{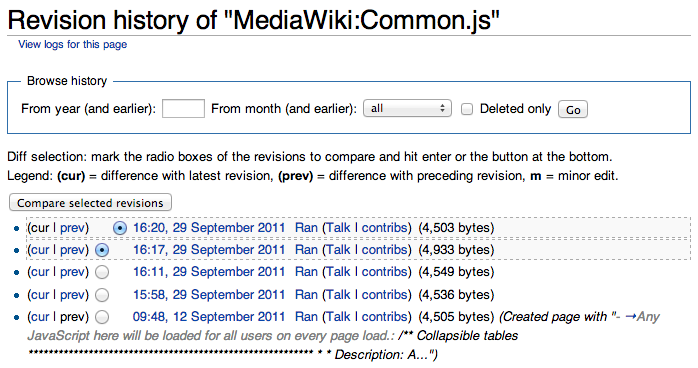}
\caption{Example of JavaScript history in MediaWiki}
\label{fig:js-history}
\end{figure}

JavaScript and StyleSheets are the other embedded resources necessary to satisfy temporal coherence.  MediaWiki natively stores all versions of stylesheets for use \cite{stylesheets}, as shown in Figure \ref{fig:css-history}.  MediaWiki also natively stores all versions of JavaScript to use \cite{javascript}, as shown in Figure \ref{fig:js-history}.

\begin{figure}[h!]
\centering
	\includegraphics[width=0.5\textwidth]{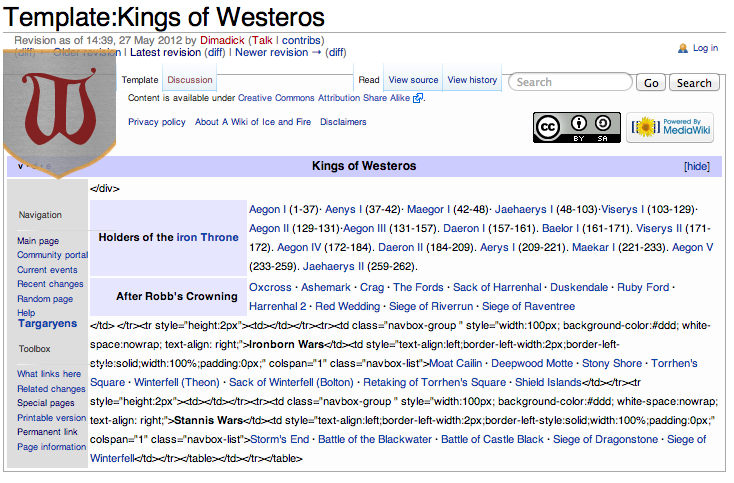}
\caption{Example of the current CSS not agreeing with an previous revision of a MediaWiki page}
\label{fig:bad-css}
\end{figure}

Figure \ref{fig:bad-css} shows an example where the CSS matters.  The previous version of this page is using the current CSS, which does not render the same way.  As a result, the shield image appears over the text on the left side of the page.

Unfortunately, we could find no hooks that allowed the MediaWiki Extension to access these resources and change how the page is rendered.  This is an item that will require us to work with the MediaWiki Development team.

Once this is achieved, it could be made an optional setting.  Some sites may not want their present content displayed with previous styles or JavaScript code.

\section{Conclusions}

We have made significant improvements to the Memento MediaWiki Extension, as identified in this paper.  The current architecture and design addresses the concerns presented by the Wikimedia community.

We have also experimented with the use of Memento Pattern 1.1 in an attempt to improve performance, and have found that it would actually have a negative impact on performance, due to idiosyncrasies in how it would need to be implemented within MediaWiki.

We have also shown how merely installing the Memento MediaWiki Extension has a negligible impact on performance for accessing MediaWiki pages, both current and oldid.

Unfortunately, until work is done with the MediaWiki development team to address embedded stylesheets and JavaScript, temporal coherence cannot be fully achieved.

There are two possible approaches.

The first approach we have already started to implement.  We have embarked on a plan of making MediaWiki perform datetime negotiation internally, generating the correct embedded resources as requested.  In this case, the Memento Protocol would only be used to acquire the correct revision of the base wiki article page, with all embedded resources changed internally to their previous states before the final response is sent back to the browser.  This is likely the best performing approach.

Alternatively, Memento, as a protocol, can be a solution here.  If MediaWiki presented all embedded resources using Memento headers, and if a TimeGate existed for each of these resources, then Memento clients could request all embedded resources the same way the original resource is requested, thus building pages from the past revisions of all resources.  This would provide a standard interface for all resources served up by MediaWiki, resulting in a cleaner MediaWiki Extension, but also requiring numerous additional requests to acquire everything needed to view a page as it existed in the past.

\section{Acknowledgments}
This work was supported in part by the Andrew Mellon Foundation.

\bibliographystyle{plain}
\bibliography{techreport}

\appendix

\begin{lstlisting}[frame=single,basicstyle=\scriptsize\ttfamily,captionpos=b,numbers=left,breaklines=true,caption={Python Source code used to process Siege Output for TimeGate (URI-G) Design Performance Testing},label={lst:timegateperfpy}, float=*]
#!/usr/bin/python

import sys
import re
from decimal import Decimal as D

def main(args):

    outputfile = args[1]

    print "outputfile:  " + outputfile

    stats = {}

    p = re.compile('.*HTTP/1.1 302[\t ]*([0-9.]*) secs:[\t ]*([0-9.]*) bytes.*$')
    p2 = re.compile('.*/([^/]*)^[\[0m$')
    p3 = re.compile('Special:TimeMap/(.*)^[\[0m$')


    f = open(outputfile)

    for line in f:

        if line[0:19] == '^[[0;36mHTTP/1.1 302':

            pagename = p2.findall(line)

            if len(pagename) == 1:
                pagename = pagename[0]
            else:
                pagename = p3.findall(line)

                if len(pagename) == 1:
                    pagename = pagename[0]
                else:
                    sys.stderr.write("FAILED TO FIND PAGE NAME\n")
                    continue

            if pagename not in stats:
                stats[pagename] = {}

            secs, bytes = p.findall(line)[0]
            n = D(secs)

            if 'demo-special' in line:
                stats[pagename]['special'] = str(n)
            elif 'demo' in line:
                stats[pagename]['default'] = str(n)

    f.close()

    print ','.join(['PAGE', 'SPECIAL', 'DEFAULT'])
    for name in stats:

        if 'special' in stats[name] and \
            'default' in stats[name]:
            print ','.join(
                [ '"' + name + '"', stats[name]['special'],
                    stats[name]['default'] ] )
        else:
            err = "Page '" + name + "' is missing statistics for "
            errlist = []
            if 'special' not in stats[name]:
                errlist.append('not-installed')
            if 'default' not in stats[name]:
                errlist.append('default')
            err += '[' + ','.join(errlist) + ']'
            sys.stderr.write(err + '\n')

if __name__ == '__main__':
    main(sys.argv)
\end{lstlisting}

\begin{lstlisting}[frame=single,basicstyle=\scriptsize\ttfamily,captionpos=b,numbers=left,breaklines=true,caption={Python Source code used to process Siege Output for URI-R and URI-M Performance Testing},label={lst:urirurimperfpy},float=*]
#!/usr/bin/python

import sys
import re
from decimal import Decimal as D

def main(args):

    inputfile = args[1]
    stats = {}

    p = re.compile('.*HTTP/1.1 200[\t ]*([0-9.]*) secs:[\t ]*([0-9.]*) bytes.*$')
    p2 = re.compile('.*title=(.*)[&]*.*^[\[0m$')
    p3 = re.compile('Special:TimeMap/(.*)^[\[0m$')

    f = open(inputfile)

    for line in f:

        if line[0:19] == '^[[0;34mHTTP/1.1 200':
            pagename = p2.findall(line)

            if len(pagename) == 1:
                pagename = pagename[0]
            else:
                pagename = p3.findall(line)

                if len(pagename) == 1:
                    pagename = pagename[0]
                else:
                    sys.stderr.write("FAILED TO FIND PAGE NAME")
                    continue

            if pagename not in stats:
                stats[pagename] = {}

            secs, bytes = p.findall(line)[0]
            n = D(secs)

            if 'demo-not-installed' in line:
                stats[pagename]['not-installed'] = str(n)
            elif 'demo-302-recommended-relations' in line:
                stats[pagename]['all-headers'] = str(n)
            elif 'demo' in line:
                stats[pagename]['default'] = str(n)

    f.close()

    print ','.join(['PAGE', 'NOT_INSTALLED', 'DEFAULT', 'ALL_HEADERS'])
    for name in stats:

        if 'not-installed' in stats[name] and \
            'default' in stats[name] and \
            'all-headers' in stats[name]:
            print ','.join(
                [ '"' + name + '"', stats[name]['not-installed'],
                    stats[name]['default'], stats[name]['all-headers'] ] )
        else:
            err = "Page '" + name + "' is missing statistics for "
            errlist = []
            if 'not-installed' not in stats[name]:
                errlist.append('not-installed')
            if 'default' not in stats[name]:
                errlist.append('default')
            if 'all-headers' not in stats[name]:
                errlist.append('all-headers')
            err += '[' + ','.join(errlist) + ']'
            sys.stderr.write(err + '\n')

if __name__ == '__main__':
    main(sys.argv)
\end{lstlisting}

\clearpage
\begin{lstlisting}[frame=single,basicstyle=\scriptsize\ttfamily,captionpos=b,numbers=left,breaklines=true,caption={Python Source code used to process Siege Output for URI-T Performance Testing},label={lst:uritperfpy},float=*]
#!/usr/bin/python

import sys
import re
from decimal import Decimal as D

def main(args):

    inputfile = args[1]
    stats = {}

    p = re.compile('.*HTTP/1.1 200[\t ]*([0-9.]*) secs:[\t ]*([0-9.]*) bytes.*$')
    p2 = re.compile('.*title=(.*)[&]*.*^[\[0m$')
    p3 = re.compile('Special:TimeMap/(.*)^[\[0m$')


    f = open(inputfile)

    for line in f:

        if line[0:19] == '^[[0;34mHTTP/1.1 200':
            pagename = p2.findall(line)

            if len(pagename) == 1:
                pagename = pagename[0]
            else:
                pagename = p3.findall(line)

                if len(pagename) == 1:
                    pagename = pagename[0]
                else:
                    sys.stderr.write("FAILED TO FIND PAGE NAME")
                    continue

            if pagename not in stats:
                stats[pagename] = {}

            secs, bytes = p.findall(line)[0]
            n = D(bytes)

            if 'demo-not-installed' in line:
                stats[pagename]['not-installed'] = str(n)
            elif 'demo-302-recommended-relations' in line:
                stats[pagename]['all-headers'] = str(n)
            elif 'demo' in line:
                stats[pagename]['default'] = str(n)

    f.close()

    print ','.join(['PAGE', 'NOT_INSTALLED', 'DEFAULT', 'ALL_HEADERS'])
    for name in stats:

        if 'not-installed' in stats[name] and \
            'default' in stats[name] and \
            'all-headers' in stats[name]:
            print ','.join(
                [ '"' + name + '"', stats[name]['not-installed'],
                    stats[name]['default'], stats[name]['all-headers'] ] )
        else:
            err = "Page '" + name + "' is missing statistics for "
            errlist = []
            if 'not-installed' not in stats[name]:
                errlist.append('not-installed')
            if 'default' not in stats[name]:
                errlist.append('default')
            if 'all-headers' not in stats[name]:
                errlist.append('all-headers')
            err += '[' + ','.join(errlist) + ']'
            sys.stderr.write(err + '\n')

if __name__ == '__main__':
    main(sys.argv)
\end{lstlisting}

\balancecolumns

\end{document}